\documentclass[11pt,aps,prd,nofootinbib,superscriptaddress,preprintnumbers]{revtex4}
\usepackage{graphicx,epsfig}
\newcommand{\bea}{\begin{eqnarray}}
\newcommand{\beq}{\begin{equation}}
\newcommand{\eea}{\end{eqnarray}}
\newcommand{\eeq}{\end{equation}}

\newcommand{\msq}{M_{\tilde{q}}}

\newcommand{\Mtlr}{\tilde M_{t}^{LR}}

\newcommand{\BR}{{\mathcal B}}

\providecommand*{\ler}{\stackrel{\scriptstyle <}{\scriptstyle \sim}}

\newcommand{\nn}{\nonumber}
\newcommand{\Frac}[2]{\frac{\displaystyle{#1}}{\displaystyle{#2}}}
\newcommand{\lsim}{\raise0.3ex\hbox{$\;<$\kern-0.75em\raise-1.1ex\hbox{$\sim\;$}}}
\newcommand{\gsim}{\raise0.3ex\hbox{$\;>$\kern-0.75em\raise-1.1ex\hbox{$\sim\;$}}}
\newcommand{\eq}[1]{Eq.~(\ref{#1})}
\newcommand{\unity}{{\hbox{1\kern-.8mm l}}}
\newcommand{\LL}{{\mbox{\scriptsize LL}}}
\newcommand{\LR}{{\mbox{\scriptsize LR}}}
\newcommand{\RL}{{\mbox{\scriptsize RL}}}
\newcommand{\RR}{{\mbox{\scriptsize RR}}}
\preprint{CPHT-RR023.0506,~~FTUV-07-0131,~~IFIC/07-04}
\begin{document}

\title{Soft SUSY Breaking Grand Unification:
 Leptons vs Quarks on the Flavor Playground}

\author{M. Ciuchini}
\affiliation{INFN, Sezione di Roma Tre and Dip. di Fisica, Univ. di Roma Tre,
Via della Vasca Navale 84, I-00146 Rome, Italy.}
\author{A. Masiero}
\affiliation{ Dip. di Fisica `G. Galilei' and
INFN, Sezione di Padova,  
Univ. di Padova, Via Marzolo 8, I-35131, Padua, Italy.} 
\author{P. Paradisi}
\affiliation{INFN, Sezione di Roma II and Dip. di Fisica, Univ. di Roma
`Tor Vergata', I-00133 Rome, Italy.}
\affiliation{Departament de F\'{\i}sica Te\`orica and IFIC, Universitat de 
Val\`encia-CSIC, E-46100, Burjassot, Spain.}
\affiliation{Department of Physics, Technion-Israel Institute of Technology,
Technion City, 32000 Haifa, Israel.}
\author{L. Silvestrini}
\affiliation{INFN, Sezione di Roma and Dip. di Fisica, Univ. di Roma
`La Sapienza', P.le A. Moro 2, I-00185 Rome, Italy.}
\author{S. K. Vempati}
\affiliation{Centre de Physique Theorique
\footnote{Unit{\'e} mixte du CNRS et de l'EP, UMR 7644.},
Ecole Polytechnique-CPHT, 91128 Palaiseau Cedex, France}
\affiliation{Centre for High Energy Physics, Indian Institute of Science,
Bangalore 560 012, India}
\author{O. Vives}
\affiliation{Departament de F\'{\i}sica Te\`orica and IFIC, Universitat de 
Val\`encia-CSIC, E-46100, Burjassot, Spain.}

\begin{abstract}
We systematically analyze the correlations between the various leptonic
and hadronic flavor violating processes arising in SUSY Grand Unified
Theories. Using the GUT-symmetric relations between the soft SUSY
breaking parameters, we assess the impact of hadronic and leptonic
flavor observables on the SUSY sources of flavor violation. 
\end{abstract}

\maketitle

\section{Introduction}

Supersymmetry (SUSY) Breaking (SB) remains one of the biggest issues
in physics beyond the Standard Model (SM). In spite of various
proposals \cite{SUSYbreaking}, we still miss a realistic and
theoretically satisfactory model of SB.

Flavor violating processes have been instrumental in guiding us
towards consistent SB models.  Indeed, even in the absence of a
well-defined SB mechanism and, hence, without a precise knowledge of
the SUSY Lagrangian at the electroweak scale, it is still possible to
make use of the Flavour Changing Neutral Current (FCNC) bounds to infer 
relevant constraints on the part of the SUSY soft breaking sector related 
to the sfermion mass matrices \cite{FCreviews}.

The model-independent method which is adopted is the so-called
Mass-Insertion Approximation (MIA) \cite{kostelesky}.  In this
approach, the experimental limits lead to upper bounds on the
parameters (or combinations of) $\delta_{ij}^f
\equiv\Delta^f_{ij}/m_{\tilde{f}}^2$, where $\Delta^f_{ij}$ is the
flavor-violating off-diagonal entry appearing in the $f = (u,d,l)$
sfermion mass matrices and $m_{\tilde{f}}^2$ is the average sfermion
mass. The mass insertions include the LL/LR/RL/RR types, according to
the chirality of the corresponding SM fermions.  Detailed bounds on
the individual $\delta$s have been derived by considering limits from
various FCNC processes \cite{gabbiani}-\cite{Ciuchini:2006dx}.

As long as one remains within the simple picture of the Minimal Supersymmetric
Standard Model (MSSM), where quarks and leptons are unrelated, the
hadronic and leptonic FCNC processes yield separate bounds on the
corresponding $\delta^q$'s and $\delta^l$'s.

The situation changes when one embeds the MSSM within a Grand Unified
Theory (GUT).  In a SUSY GUT, quarks and leptons sit in the same
multiplets and are transformed into each other through GUT symmetry
transformations.  In supergravity theories, where the supersymmetry
breaking (SB) mediation to the visible sector is gravitational, the
structure of the soft terms is essentially dictated by the K\"ahler
potential.  If the effective supergravity Lagrangian is defined at a
scale higher than the Grand Unification scale, the matter fields
present in the K\"ahler function have to respect the underlying gauge
symmetry which is the GUT group itself. Subsequently, the SB mediation
would give rise to the usual soft terms which however now follow the
GUT symmetry.  Hence, we expect quark-lepton correlations among
entries of the sfermion mass matrices \cite{ourprl,othercorrelations}.
In other words, the quark-lepton unification seeps also into the SUSY
breaking soft sector.

Imposition of a GUT symmetry on the soft SUSY breaking Lagrangian
$\mathcal{L}_{\rm soft}$ entails relevant implications at the weak
scale. This is because the flavor violating (FV) mass-insertions do
not get strongly renormalized through Renormalization Group (RG)
scaling from the GUT scale to the weak scale in the absence of new
sources of flavor violation. On the other hand, if such new sources
are present, for instance due to the presence of new neutrino Yukawa
couplings in SUSY GUTs with a seesaw mechanism for neutrino masses,
then one can compute the RG-induced effects in terms of these new
parameters. As has been noted earlier \cite{ourprl}, even in such
cases, the correlations between hadronic and leptonic flavor violating
MIs survive at the weak scale as a function of these parameters.  As
for the flavor conserving (FC) mass insertions (i.e., the diagonal
entries of the sfermion mass matrices), they get strongly renormalized
but in a way which is RG computable.

In conclusion, in SUSY GUTs where the soft SUSY breaking terms respect
boundary conditions which are subject to the GUT symmetry to start
with, we generally expect the presence of relations among the 
(bilinear and trilinear) scalar terms in the hadronic and leptonic
sectors. Such relations hold true at the (superlarge) energy scale
where the correct symmetry of the theory is the GUT symmetry. After its
breaking, the mentioned relations will undergo corrections which are
computable through the appropriate RGE's which are related to the
specific structure of the theory between the GUT and the electroweak
scale (for instance, new Yukawa couplings due to the presence of
right-handed (RH) neutrinos acting down to the RH neutrino mass scale,
presence of a symmetry breaking chain with the appearance of new
symmetries at intermediate scales, etc.). As a result of such a
computable running, we can infer the correlations between the softly
SUSY breaking hadronic and leptonic $\delta$ terms at the low scale
where we perform our FCNC tests. Explicit examples of such
correlations in the context of an SU(5) SUSY GUT will be provided in
next Section.

Given that a common SUSY soft breaking scalar term of
$\mathcal{L}_{\rm soft}$ at scales close to $M_{\rm Planck}$ can give rise to
RG-induced $\delta^q$'s and $\delta^l$'s at the weak scale, one may
envisage the possibility to make use of the FCNC constraints on such
low-energy $\delta$'s to infer bounds on the soft breaking parameters
of the original supergravity Lagrangian
($\mathcal{L}_{\rm sugra}$). Indeed, for each scalar soft parameter of
$\mathcal{L}_{\rm sugra}$ one can ascertain whether the hadronic or the
corresponding leptonic bound at the weak scale yields the stronger
constraint at the large scale. This constitutes the major goal of this
work: we intend to go through an exhaustive list of the low-energy
constraints on the various $\delta^q$'s and $\delta^l$'s and, then,
after RG evolving such $\delta$'s up to $M_{\rm Planck}$, we will
establish for each $\delta$ of $\mathcal{L}_{\rm sugra}$ which one between
the hadronic and leptonic constraints is going to win, namely which
provides the strongest constraint on the corresponding
$\delta_{\rm sugra}$.

We will show that there exists a very interesting complementarity in
the sensitivity of the soft breaking sector of $\mathcal{L}_{\rm sugra}$
to the FCNC constraints provided by hadronic and leptonic physics.

The second, related purpose of this paper is to fully exploit the
common origin from single $\delta_{\rm sugra}$'s of hadronic and
leptonic weak scale $\delta$'s to make use of leptonic (hadronic) FCNC
constraints to limit $\delta^q$'s ($\delta^l$'s). In other words, for
instance in a SUSY SU(5) context, one can use a leptonic FCNC process
like $\tau \to \mu \gamma$ to impose constraints on the
$\left(\delta^d_{23}\right)_{\RR}$ term of the hadronic sector which
are stronger than those which are derived from genuine hadronic FCNC
processes like $b \to s\gamma$ \cite{ourprl}.
 
 In this respect, the present work intends to provide a particularly
 striking example of the correlation between ``low-energy'' (i.e. weak
 scale) experiments and ``high-energy'' (i.e. GUT or Planck scale) SUSY
 Lagrangian. The effort in the coming years, provided LHC yields some
 SUSY evidence, will be to ``reconstruct'' the original supergravity
 Lagrangian from which our weak scale testable SUSY descends. As we
 know, such work of reconstruction will be rather painful if we have
 to rely only on LHC results; on the other hand, accompanying the SUSY
 direct searches at LHC with the powerful FCNC tests will prove to be
 quite efficient in our effort of tracing back the original parameters
 entering the underlying supergravity Lagrangian. In the case of SUSY
 GUTs such role of FCNC processes is further enhanced because of the
 above mentioned hadron--lepton correlations.
 
 The paper is organized as follows. In the next Section we will
 provide an example of relations between hadronic and leptonic FC
 $\delta's$ at the weak scale within SUSY SU(5). 
In Section \ref{sec:basis}, we discuss the impact of GUT breaking
effects (both at the tree level and at 1-loop) and the subsequent
basis dependence which seeps in to the GUT-symmetric relations. 
 In Section \ref{sec:Constraints}, we discuss our procedure for
obtaining bounds detailing various constraints we have imposed
on our supersymmetric spectrum. 
In Sections \ref{sec:MIhadronic} and
 \ref{sec:leptons}, we proceed with a separate analysis of the hadronic
 and leptonic $\delta's$, to conclude with a study of their
 correlations in Section \ref{sec:correlations}. In Section
 \ref{sec:conclusions}, the conclusions include some kind of final
 ``score'' on the tightness of the bounds in GUT related $\delta's$
 coming from the various FCNC processes and the outlook for the
 interplay of the kind of flavor physics we consider here with the
 possible LHC outcome.

\section{On hadron--lepton FCNC relations in SUSY GUTs}
\label{sec:GUTrel}
In this Section we provide an example of the correlations
 between hadronic and leptonic $\delta's$ entering the weak scale MSSM
 Lagrangian when the underlying theory at the large scale, where
 supergravity lives, is restricted by a grand unified symmetry.  Let
 us consider the scalar soft breaking sector of the MSSM:
 
\bea 
\label{smsoft}
- {\cal L}_{\rm soft} &=& m_{Q_{ii}}^2 \tilde{Q}_i^\dagger \tilde{Q}_i 
+ m_{u^c_{ii}}^2 \tilde{u^c}_i^\star \tilde{u^c}_i + m^2_{e^c_{ii}} 
\tilde{e^c}_i^\star \tilde{e^c}_i 
 +   m^2_{d^c_{ii}} \tilde{d^c}^\star_i 
\tilde{d^c}_i + m_{L_{ii}}^2 \tilde{L}_i^\dagger \tilde{L}_i + 
m^2_{H_1} H^\dagger_1 H_1 
+  m^2_{H_2} H_2^\dagger H_2  \nonumber \\ &+& A^u_{ij}~
\tilde{Q}_i \tilde{u^c}_j H_2 + A^d_{ij}~
\tilde{Q}_i \tilde{d^c}_j H_1 + A^e_{ij}~
\tilde{L}_i \tilde{e^c}_j H_1 + 
(\Delta^l_{ij})_{\LL} \tilde{L}_i^\dagger \tilde{L}_j  + 
(\Delta^e_{ij})_{\RR} \tilde{e^c}_i^\star \tilde{e^c}_j  \nonumber \\ 
&+& (\Delta^q_{ij})_{\LL} \tilde{Q}_i^\dagger \tilde{Q}_j  + 
(\Delta^u_{ij})_{\RR} \tilde{u^c}_i^\star \tilde{u^c}_j  + 
(\Delta^d_{ij})_{\RR} \tilde{d^c}_i^\star \tilde{d^c}_j  
+ (\Delta^e_{ij})_{\LR} \tilde{e_L}_i^\star \tilde{e^c}_j  
+ (\Delta^u_{ij})_{\LR} \tilde{u_L}_i^\star \tilde{u^c}_j \nonumber \\ 
&+& (\Delta^d_{ij})_{\LR} \tilde{d_L}_i^\star \tilde{d^c}_j + \ldots 
\eea
where we have used the standard notation for the MSSM fields and have 
explicitly written down the various $\Delta$ parameters. Notice that 
the different $\Delta_\LR$ include the contributions from the trilinear 
terms with the corresponding Higgs vev. Therefore the trilinear couplings 
in the second line represent only the real Higgs couplings and not their 
vacuum expectation values.

Consider $SU(5)$ to be the relevant symmetry at the scale where the
soft terms are generated.  Then, taking into account that
matter is organized into the SU(5) representations ${\bf 10}~
=~(q,u^c,e^c)$ and ${\bf \overline 5}~ = ~(l,d^c)$, one obtains the
following relations
\bea
\label{matrel1}
m^2_{Q} = m^2_{\tilde{e^c}} = m^2_{\tilde{u^c}} = m^2_{\bf 10} \\
\label{matrel2}
m^2_{\tilde{d^c}} = m^2_{L} = m^2_{\bar{\bf \overline 5}} \\
\label{trirel}
A^e_{ij} = A^d_{ji}\, .
\eea
Eqs.~(\ref{matrel1})--(\ref{trirel}) are matrices in flavor
space.  These equations lead to relations between the slepton and squark 
flavor violating off-diagonal entries $\Delta_{ij}$. These are: 
\bea
\label{cdeltas1}
(\Delta^u_{ij})_{\LL} = (\Delta^u_{ij})_{\RR} = (\Delta^d_{ij})_{\LL} =
(\Delta^l_{ij})_{\RR} \\
\label{cdeltas3}
(\Delta^d_{ij})_{\RR} = (\Delta^l_{ij})_{\LL} \\
\label{cdeltas4}
(\Delta^d_{ij})_{\LR} = (\Delta^l_{ji})_{\LR} = (\Delta^l_{ij})_{\RL}^\star
\eea
These GUT correlations among hadronic and leptonic scalar soft terms
are summarized in the second column of Table~\ref{tb0}. Assuming that
no new sources of flavor structure are present from the $SU(5)$ scale
down to the electroweak scale, apart from the usual SM CKM one, one
infers the relations in the first column of Table~\ref{tb0} at the
low-energy scale.  
 \begingroup
 \begin{table}
 \begin{center}
 \begin{tabular*}{0.8\textwidth}{@{\extracolsep{\fill}}||c|c||}
 \hline\hline
 Relations at the weak scale & Relations at $M_{\rm GUT}$ \\[0.2pt] 
 \hline
  $(\delta^u_{ij})_{\RR}~ \approx~ (m_{e^c}^2/ m_{u^c}^2)~ 
 (\delta^l_{ij})_{\RR}$ & $m^2_{{u^c}_0} ~=~ m^2_{{e^c}_0}$ \\
 \hline
  $(\delta^q_{ij})_{\LL}~ \approx~(m_{e^c}^2/ m_{Q}^2)~ (\delta^l_{ij})_{\RR}$ &
 $m^2_{{Q}_0} ~=~ m^2_{{e^c}_0}$ \\ 
 \hline 
 $(\delta^d_{ij})_{\RR} ~\approx~ (m_{L}^2/ m_{d^c}^2)~ (\delta^l_{ij})_{\LL}$ &
 $m^2_{{d^c}_0} ~=~ m^2_{{L}_0}$ \\ 
 \hline 
 $(\delta^d_{ij})_{\LR}~\approx~ (m_{L_{avg}}^2 /m_{Q_{avg}}^2) ~
 (m_b/ m_\tau) ~ (\delta^l_{ij})_{\LR}^\star$  & 
 $A^e_{{ij}_0} = A^d_{{ji}_0}$
\\\hline\hline
 \end{tabular*}
 \end{center}
 \caption{Links between various transitions between up-type, down-type quarks 
 and charged leptons for SU(5). The suffix `0' implies GUT scale parameters.}
 \label{tb0}
 \end{table}
 \endgroup
A comment is in order when looking at Table~\ref{tb0}. 
Mass insertions for down-type quarks and leptons in Table~\ref{tb0}
always exhibit opposite ``chiralities'', i.e. LL insertions are 
related to RR ones and vice-versa.  This stems from the arrangement 
of the different fermion chiralities in $SU(5)$ multiplets
(as it clearly appears from the last column in Table~\ref{tb0}).
This restriction can easily be overcome if we move from $SU(5)$ to
left-right symmetric unified models like SO(10) or the Pati-Salam (PS)
case,
with gauge group $SU(4)_c \times SU(2)_L \times SU(2)_R$.

\section{Weak-scale quark-lepton relations}
\label{sec:basis}

The exact equality of the mass and trilinear matrices of the different
components of a GUT multiplet is only true when the GUT symmetry is valid.
At low scales, where we perform our experiments, these relations are modified.
In this section, we analyze how these relations are modified after the 
breaking of the GUT symmetry and the RGE running from $M_{\rm GUT}$ to $M_W$. 

\subsection{Definition of the SCKM basis}

In low-energy phenomenology we obtain the different mass insertion bounds 
in the so-called super-CKM (SCKM) basis, i.e. the basis where through 
a  rotation of the whole superfield (fermion + sfermion) we obtain diagonal
Yukawa couplings for the corresponding fermion field \cite{kingkane}. At low
energies the 
SCKM  bases  for quarks and leptons are unrelated. However in a GUT theory
quarks and leptons unify and thus we would expect the quark and lepton Yukawa
couplings to be equal  at the GUT scale and hence their SCKM bases to unify.  
Unfortunately this is not true 
when we evolve the Yukawa couplings using the SM (or MSSM)
Renormalization Group Equations (RGEs). This discrepancy, that could be solved
with some GUT breaking effects or new non-renormalizable contributions,
implies a relative rotation between the quark and leptonic SCKM
basis. While in principle this is an obstacle for a model-independent
analysis, we argue below that its effects are not expected to change the
order of magnitude of our bounds.

Let us consider for example the case in which we obtain a MI bound
from a leptonic process at low energies. This bound is given in the
basis of diagonal charged lepton masses. The next step in our scheme
consists in using the RGE equations to evolve this bound to the GUT
scale. The resultant bound is a bound on the element in the sfermion
mass matrix common for the GUT multiplet including both quark and
leptons at $M_{\rm GUT}$ in the basis of {\em diagonal charged lepton
  Yukawa matrices}.\footnote{An additional (small) RGE-induced
  rotation can be necessary in the presence of neutrino Yukawa
  couplings. We neglect its effect in our discussion.} However, if we
want to convert this bound into a bound on the corresponding squark MI
we must take into account a possible misalignment between the
charged-lepton and the down-quark Yukawa matrices just below the GUT
scale. GUT symmetry breaking effects could introduce different
corrections to the quark and lepton Yukawa matrices. For example, it
is well-known that the minimal $SU(5)$ Lagrangian with a ${\bf 5}$ and
a ${\bf \bar 5}$ Higgs representations is not realistic as the first
two generations of the down quark and charged lepton fermion masses do
not follow the GUT relations. This means that quark and lepton Yukawa
matrices are not the same below $M_{\rm GUT}$ and they are not
simultaneously diagonalizable.  Taking into account these facts, we
must rotate our bounds from the leptonic sector to the basis of
diagonal quark Yukawa matrices before we compare them with the bounds
obtained from quark processes at low energies.

In a complete GUT model we would be able to obtain both the quark and
lepton Yukawa matrices after the breaking of the GUT symmetry.
In such a theory it would be possible to compute the 
relative rotation between the basis of diagonal quark and
the basis of diagonal lepton Yukawa matrices exactly. On more general 
grounds, as we do not have a complete GUT theory yet, we have a new
uncertainty introduced in the translation of the bounds between quark 
and leptonic MIs, due to this relative rotation between the leptonic 
and quark Yukawa couplings. To see the effect of this 
`new' rotation matrix, which can be represented by a unitary matrix,
$V^{(ql)}$, while translating the bounds from one sector to another
sector (leptonic to hadronic or vice-versa), we consider one particular
case. Let $\Delta^l_{i\neq j}$, of any `chirality', have a bound at the
GUT scale, given by $\Delta_{max}$. 
This element $\Delta^l_{i\neq j}$ corresponds to a combination of matrix
elements in the basis of diagonal quark Yukawas, $\Delta^{(q)}_{i j}$,
given as
\bea 
|\Delta^{(l)}_{i\neq j}| = |\sum_{k,l} V^{(ql)\,*}_{k i}
\Delta^{(q)}_{k l} V^{(ql)}_{l j}| \leq \Delta^{max}, 
\eea 
where $V^{(ql)}$ would represent the rotation between the basis of
diagonal quark and lepton Yukawas. Given that $\Delta^{(f)}$ and $V^{(ql)}$ 
have different origins, we do not expect large cancellations between 
different terms in this sum and therefore, {\it
``barring accidental cancellations''}, we can apply the bound on
$\Delta^{(l)}_{i\neq j}$ to each of the individual elements in
this sum  
\bea 
\label{fulltrans}
\left |V^{(ql)\,*}_{k i}~\Delta^{(q)}_{k l}~ V^{(ql)}_{l j}\right|~ \leq ~
\Delta^{max}, 
\eea
Once more, in principle it is necessary to know the relative rotation
$V^{(ql)}$ to apply the bound on $\Delta^{(q)}_{k l}$. In SU(5) GUT, 
requiring that the successful
$b$--$\tau$ unification is not accidental, we can expect that 
$|V^{(ql)}_{3 3}| \simeq 1$ and $|V^{(ql)}_{3\neq i}|,|V^{(ql)}_{i\neq3}|\ll
1$. The relative misalignment for the first two generations is much more
uncertain. However, we consider reasonable to expect that the different
families are approximately arranged in the usual way and $(c,s)$ quarks are
mostly associated with $(\mu,\nu_\mu)$ and $(u,d)$ with $(e,\nu_e)$. So, we
make the ``reasonable assumption'' that  diagonal elements in these rotations, 
$V^{(ql)}_{i i}$, are ${\cal{O}}(1)$, where ${\cal{O}}(1)$ may mean
$1/\sqrt{2}$ or even $1/2$ but not much smaller. Therefore, under these
conditions we have that 
\bea
\label{approxrot} 
\left |V^{(ql)\,*}_{i i}~\Delta^{(q)}_{i\neq j}~ V^{(ql)}_{j j}\right|~ \simeq
~\left|\Delta^{(q)}_{i j}\right|~ \leq ~\Delta^{max}, 
\eea
and thus, {\em order of magnitude MI bounds} in the leptonic sector can be
safely translated into {\em order of magnitude MI bounds} in the quark sector,
taking into account that these bounds are not precise within factors of two.

In the following we present a realistic example to show the possible 
effects of these rotations. A popular solution to the first two generations
 Yukawa unification problem in SU(5) is provided by the Georgi-Jarlskog 
mechanism \cite{GeorgiJarls}.  
There one introduces an additional Higgs representation which is \textbf{45}
dimensional contributing mainly to the 
second generation Yukawa couplings. This Higgs gets a vev along the 
$(B - L)$ direction and therefore 
breaks the charged lepton-down quark symmetry. Assuming simple
symmetric Yukawa textures, the down quark and charged lepton Yukawa matrices 
are given as follows: 
\beq 
Y_d^{M_{\rm GUT}} \propto
\left( \begin{array}{ccc} 0 & \lambda^3 & \lambda^3 \\ \lambda^3 &
\lambda^2 & \lambda^2 \\ \lambda^3 & \lambda^2 & 1
\end{array} \right) \qquad \qquad Y_e^{M_{\rm GUT}} \propto
\left( \begin{array}{ccc} 0 & \lambda^3 & \lambda^3 \\ \lambda^3~ &
~3\lambda^2 &~ \lambda^2 \\ \lambda^3 & \lambda^2 & 1
\end{array} \right)\,,
\eeq
where $\lambda$ is a parameter of the order of the Cabibbo angle.
In this way, we preserve the successful $b$--$\tau$ unification and correct 
the bad GUT relations for the second and first generations. Clearly, if the 
Yukawa matrices have this structure, the basis of diagonal down-quark 
Yukawas is different from the basis of diagonal charged-lepton Yukawas. 
Therefore the rotation $V^{(ql)}$, due to the misalignment between quarks 
and leptons is non-trivial. However, it is very easy to check that up to 
terms of $O(\lambda^3)$ this rotation is given by
\beq 
V^{(ql)} \simeq
\left( \begin{array}{ccc} 1 - \Frac{2 \lambda^2}{9} & \Frac{2 \lambda}{3}& 
0 \\ - \Frac{2 \lambda}{3} &1 - \Frac{2 \lambda^2}{9} & 0 \\ 0 & 0 & 1
\end{array} \right).
\eeq
Using this rotation and barring accidental cancellations in \eq{fulltrans}, 
it is evident that 
\eq{approxrot} is an excellent approximation in this case and we can safely
relate off-diagonal elements in the squark and slepton mass matrices.

\subsection{RG evolution}
\label{sec:RGE}

The SU(5) GUT relations between squark and slepton masses and
trilinear couplings in Eqs.~(\ref{matrel1})-(\ref{cdeltas4}) are only
valid at the SU(5) scale. As it is well-known, one could expect the
long RG running from the GUT scale down to the weak scale to modify
these relations significantly.  However, the crucial aspect is that
the flavor violating entries ($i \neq j$) are not significantly
modified due to this running.  In
fact, even in the presence of right handed neutrinos in a 
seesaw mechanism, this statement remains true up to factors order one. This
approximate non-modification of the $\Delta_{i\neq j}$'s,
allows us to recast them at the weak scale in terms of $\delta_{i\neq j}$. 
In the present section, we will try to quantify our statements
using semi-numerical solutions of the RGE within the SU(5) model we
have been discussing so far. We will consider the effects of adding
right-handed neutrinos a bit later on.

In the RG evolution of diagonal elements of sfermion mass matrices we neglect
small Yukawa couplings and keep only $Y_b$, $Y_\tau$ and $Y_t$. Then,
for the first  
two generations of squarks and sleptons, the following simple
expressions hold in terms of the high scale soft parameters, 
$m^2_{\bf 10} = m^2_{\bf \bar 5} = m_0^2$ 
and $M_{1/2}$ (for a general expression with different $m_{\bf 10}^2$ and 
$m_{\bf \bar 5}^2$ see Table I in \cite{oscar01}) :
\bea
(m^2_{Q})_{1,2}(M_W) &\simeq& m_{0}^2 + 6.5~ M^2_{1/2}\qquad
(m^2_{D})_{1,2}(M_W) \simeq m_{0}^2 + 6.1~ M^2_{1/2} \nn\\
(m^2_{E})_{1,2} (M_W)&\simeq& m_{0}^2 + 0.15~ M^2_{1/2}\qquad
(m^2_{L})_{1,2} (M_W)\simeq m_{0}^2 + 0.5~ M^2_{1/2} \nn \\
(m^2_{U})_{1,2}(M_W) &\simeq& m_{0}^2 + 6.2~ M^2_{1/2}
\eea
Third generation masses receive contributions proportional to Yukawa
couplings which depend on the value of $\tan \beta$.
For low $\tan \beta\simeq 5$ we have
\bea
(m^2_{Q})_{3}(M_W) &\simeq& 0.6~m_0^2 + 5.5~ M^2_{1/2}\qquad
(m^2_{D})_{3}(M_W) \simeq m_0^2 + 5.8~ M^2_{1/2} \nn\\
(m^2_{E})_{3} (M_W)&\simeq& m_0^2 + 0.16~ M^2_{1/2}\qquad
(m^2_{L})_{3} (M_W)\simeq m_0^2 + 0.5~ M^2_{1/2} \nn \\
(m^2_{U})_{3}(M_W) &\simeq& 0.2~m_0^2 + 4.1~ M^2_{1/2}\,,
\eea
while for $\tan \beta\simeq 30$ we have
\bea
(m^2_{Q})_{3}(M_W) &\simeq& 0.5~m_0^2 + 5.2~ M^2_{1/2}\qquad
(m^2_{D})_{3}(M_W) \simeq 0.8~m_0^2 + 5.14~ M^2_{1/2} \nn\\
(m^2_{E})_{3} (M_W)&\simeq& 0.8~ m_0^2 +  0.12~ M^2_{1/2}\qquad
(m^2_{L})_{3} (M_W)\simeq 0.92 m_0^2 +  0.5 M^2_{1/2} \nn \\
(m^2_{U})_{3}(M_W) &\simeq& 0.2~m_0^2 + 4.2~ M^2_{1/2}\,.
\eea
RGE evolution for off-diagonal elements is solely
dependent on Yukawa couplings and does not include gaugino contributions.
Now, if we neglect first two generations Yukawa couplings,in the basis of
diagonal down-quark Yukawa couplings we have:
\bea
\label{rgeMoD}
(16\pi^2){d (m^2_{D,E})_{i\neq j} \over dt}& =& - 2 \,Y_{b,\tau}^2\, 
(m^2_{D,E})_{3j}\,\delta_{i3}\, - 2 \,Y_{b,\tau}^2\, 
(m^2_{D,E})_{i3}\,\delta_{j3} \nn \\
(16\pi^2){d (m^2_{Q})_{i\neq j} \over dt}& =& - \left[ 
Y_{b}^2\, (m^2_{Q})_{3j}\,\delta_{i3}\, + \,Y_{b}^2\, 
(m^2_{Q})_{i3}\,\delta_{j3}\, + \,Y_{t}^2\,V_{t i}^* V_{t k}\,   
(m^2_{Q})_{kj}\, + \,\right. \nn \\ && \left. Y_{t}^2\, V_{t k}^* V_{t j}\, 
(m^2_{Q})_{ik}\, +  2 \,Y_{t}^2\, V_{t i}^* V_{t j}\, 
\left( (m^2_U)_{33}\, + \, m^2_{H_u}\right) \, + \, 2 \, \left( Y^u_A
  Y_A^{u\,\dagger} \right)_{ij}\right] \nn \\
(16\pi^2){d (m^2_{L})_{i\neq j} \over dt} &=& - 
Y_{\tau}^2\, (m^2_{L})_{3j}\,\delta_{i3}\, - \,Y_{\tau}^2\, 
(m^2_{L})_{i3}\,\delta_{j3} 
\eea
where $t = \log M_{\rm GUT}/Q$. We are interested in two different aspects of
these equations. First we want to determine how a non-vanishing off-diagonal 
entry in the sfermion mass matrices at $M_{\rm GUT}$ is modified due to RGE
evolution from $M_{\rm GUT}$ to $M_W$. Then, we are also interested in the size
of the off-diagonal entries generated by the running from $M_{\rm GUT}$ to
$M_W$ in the case of exactly vanishing off-diagonal entries at $M_{\rm GUT}$. 

Regarding the evolution of GUT elements, we can neglect the non-diagonal CKM
elements and then, we see from \eq{rgeMoD} that the 
$(\Delta^{\tilde{f}})_{12,21}$ are not modified at the leading order. 
Furthermore, in the low tan$\beta$ limit, off-diagonal elements in
$m^2_{D}$ or $m^2_{E}$ in the SCKM basis are not significantly changed 
through RG evolution. Thus, one can safely assume 
Eq.~(\ref{cdeltas3}) to be valid at any scale at low tan$\beta$.  
For large $\tan \beta \simeq 40$, the effect of $Y_{b,\tau}^2$ is only 
relatively important and in the leading log approximation 
we obtain
\beq
\label{leading1}
(m^2_{D})_{i=3\neq j} (M_W) 
\simeq 
(m^2_{L})_{i=3\neq j} (M_W)~ \mbox{Exp.} \left(- \frac{1}{16 \pi^2} (2 Y_{b}^2 -
Y_{\tau}^2) \log{\frac{M_{\rm GUT}}{M_W}} \right) \simeq 0.84~(m^2_{L})_{i=3\neq j}
(M_W)
\eeq
and therefore it can account at most for a (15-20)\% relative change. 
Something analogous happens in the left-handed squark mass matrix. 
For low $\tan \beta$ the effect of the top Yukawa coupling in the third row 
elements would be
\bea
\label{leading2}
(m^2_{Q})_{i=3\neq j} (M_W) \simeq (m^2_{E})_{i=3\neq j} (M_W) \, \mbox{Exp.}
\left(- \frac{1}{16 \pi^2} Y_t^2 \log{\frac{M_{\rm GUT}}{M_W}} \right) \simeq  
0.86 (m^2_{E})_{i=3\neq j} (M_W),
\eea
In the large $\tan \beta$ region, we have to multiply the argument in the
exponential by a factor $(Y_t^2 + Y_b^2 - 2 Y_\tau^2)/Y_t^2$, which for 
$\tan \beta = 40$ amounts only to a factor of $1.16$. Therefore, the 
relative change between $(m^2_{Q})_{i=3\neq j}$ and $(m^2_{E})_{i=3\neq j}$ 
is practically the same as the previous case. 

A second feature of these RGE equations is that, even if we start from a 
flavor blind situation at the GUT (or Plank) scale, flavor off-diagonal 
entries in the LL squark mass matrix at the weak scale (via the CKM matrix)
are unavoidable. The running from $M_{GUT}$ to the weak scale $M_W$ gives rise
to the following FV effects
\begin{eqnarray}
\label{dllckm}
(\delta_{LL}^d)_{ij}\;\simeq\; 
-\frac{1}{8\pi^2} Y_t^2 V^{*}_{ti}V_{tj} 
\frac{3m_0^2 + a_0^2}{m_{\tilde q}^2} \ln \frac{M_{\rm GUT}}{M_W}.
\label{dLL}
\end{eqnarray}
If we take $M_{\rm GUT} \simeq 2 \times 10^{16}$ GeV, $Y_t\simeq 1$, $m_{\tilde q}^2\simeq 6
M_{1/2}^2 + m_0^2$ and in the limit $M_{1/2}\simeq m_0\simeq a_0$, we have  
$(\delta_{LL}^d)_{ij} \simeq -0.2 ~V^{*}_{ti}V_{tj}$. More generally we can
say that $(\delta_{LL}^d)_{ij} \simeq - c  ~V^{*}_{ti}V_{tj}$ with $c$ a
numerical coefficient depending on $(m_0,M_{1/2}, a_0)$ taking values between
$0.1$ and 1. 
As we will discuss in following sections, these RGE induced FV effects
have an important phenomenological impact, specially when FV entries
in the RR squark sector are also present.
This is due to the fact that, as we will see, $\Delta M_{B,K}$ mass 
differences are much more sensitive to contributions proportional to 
$\delta_{\rm  LL}\times\delta_{\rm RR}$.

Finally, for the last equation in (\ref{cdeltas4}),
the RGE equations for off-diagonal down quark and charged lepton trilinear 
couplings are 
\begin{eqnarray}
\label{rgetri}
(16\pi^2){d (Y^A_d)_{i\neq j} \over dt}& =& \left(
\frac{16}{3} g^2_3 \,+ \, 3 g_2^2 \,+ \,\frac{7}{9} g_1^2 \right)
(Y^A_d)_{ij}\, - \, 5   Y_b^2\, (Y^A_d)_{3j}\,\delta_{i3}\, - 
\, 4  Y_b^2\, (Y^A_d)_{i3}\, \delta_{j3} \\
&&- 3 Y_b^2 (Y^A_d)_{ij}\, - \, Y_t^2 \,(Y^A_d)_{3j}\, \delta_{i3} \, 
- \, Y_\tau^2\, (Y^A_d)_{ij} 
\nonumber \\
(16\pi^2){d (Y^A_e)_{i\neq j} \over dt} &=& \left(
3 g_2^2 \,+ \,3 g_1^2 \right)
(Y^A_e)_{ij}\, - \, 5  Y_\tau^2\, (Y^A_e)_{3j}\, \delta_{i3}\, - 
\, 4  Y_\tau^2\, (Y^A_e)_{i3}\, \delta_{j3}\, - \, Y_\tau^2\, (Y^A_e)_{ij} 
\nonumber \\
&&- 3 Y_b^2\, (Y^A_e)_{ij}. \nn
\end{eqnarray}
From the equations above, it's clear that the trilinear couplings scale
almost in the same manner as the corresponding Yukawa couplings. Thus
the corresponding relation given in Table I holds well especially in
the limit of small tan$\beta~ \ler~25$, where we can neglect the bottom
and $\tau$ Yukawas.  For large tan$\beta$ and for
$13$ and $23$ MI's, we expect similar corrections $\sim~(15-20)\%$ as
we have mentioned above.  Given that the MI bounds are intrinsically
order of magnitude bounds, we can safely consider these off-diagonal
elements to evolve analogously to the Yukawa couplings from $M_W$ to
$M_{\rm GUT}$.

Let us now briefly mention the implication of adding right-handed
neutrinos as in a seesaw mechanism. The RG effects and subsequent 
flavour violation have been studied in detail in the literature
\cite{fbam,ourreview}. The implication of these effects on the 
relations Eqs.~(\ref{cdeltas1}-\ref{cdeltas4}) depends crucially on 
the strength of the neutrino Dirac yukawa couplings $Y_\nu$ and 
the ``mixing'' they carry. For example, a rough estimate of the weak scale 
$(\Delta_{LL}^l)_{ij}$ is given by 
\beq
\label{estimate}
(m_{\tilde{L}})^2_{i\neq j} (M_W) \approx (m_{\tilde{L}})^2_{i\neq j}
(M_{GUT}) - \frac{1}{8 \pi^2}\sum_{k} (Y_\nu Y_\nu^{\dagger})_{ij} (3
m_0^2 + A_0^2) \log \left({M_{GUT} \over M_{R_k}}\right) \eeq where
$Y_\nu$ represents the neutrino Dirac Yukawa couplings. From the
equation above we see that RG effects could ``generate'' large
$\Delta$s if the mixing in $Y_\nu$ is large.  In the general case, it
is very difficult to separate the RGE-induced $\Delta$s from the
RGE-modified GUT off-diagonal elements in $m_{\tilde L}^2$ and
$Y^A_e$.  Changes in the original GUT off-diagonal entries and the
RGE-induced $\Delta$s depend on the neutrino Yukawa couplings. These
neutrino Yukawa matrices are present in the RGE between the scales
$M_{\rm GUT}$ and the mass of the corresponding right handed neutrino.
Typically one of the eigenvalues in the neutrino Yukawa matrices can
be of the same order as the top Yukawa coupling,\footnote{We do not
  consider here the possibility of having several large eigenvalues in
  $Y_{\nu}$.} but this large eigenvalue is necessarily associated with
the heaviest right-handed neutrino \cite{understanding,so10} and
therefore it decouples early. The main problem regarding the neutrino
Yukawa couplings is that the mixing diagonalizing $Y_\nu
Y_\nu^\dagger$ in the basis of diagonal $Y_e Y_e^\dagger$ is unknown
and could be large.  Nevertheless, we can discuss some cases.

The simplest situation corresponds to the case where the mixing
diagonalizing the charged lepton and neutrino Yukawa matrices are
both small. In this case, the large neutrino mixings are generated
through the seesaw mechanism and they play no role in the RGE
evolution of slepton matrices. The RGE evolution of $m^2_{\tilde L}$ is then
very similar to the previous cases without the right handed neutrinos. 
In the low $\tan \beta$ region
we can completely neglect the effects of $Y_e Y_e^\dagger$ in the
RGEs. Then in the basis of diagonal neutrino Yukawas
(approximately equal to the basis of diagonal charged-lepton
Yukawas) the off-diagonal elements would be 
\beq 
\label{leadingnu}
(m^2_{D})_{i=3\neq j} (M_W) \simeq (m^2_{L})_{i=3\neq j} (M_W)~ 
\mbox{Exp.} \left( \frac{1}{16 \pi^2} (Y_{\nu_3}^2)
\log{\frac{M_{\rm GUT}}{M_{R_3}}} \right) \simeq
1.1~(m^2_{L})_{i=3\neq j} (M_W), 
\eeq 
taking $M_{R_3}$ as low as $10^{10}$ GeV.

In the large $\tan \beta$ region, the effects of
\eq{leadingnu} and \eq{leading1} have different signs and partially
cancel out.  So, we can also expect these off-diagonal
elements to remain of similar magnitude between $M_{\rm GUT}$ and
$M_W$. Notice also that the $(1,2)$ mass insertions remain
unchanged if we have only one large neutrino-Yukawa coupling
corresponding to the third generation.

The second situation corresponds to large mixings in the neutrino
Yukawa matrices in the basis of diagonal charged-lepton Yukawas. This
case depends strongly on the mixings and size of the neutrino Yukawa
couplings.  There are two simultaneous effects, the creation of a new
MI due to the RGE and the change from RG evolution of the GUT delta.
These effects cannot be sorted out in general and thus we cannot
describe the evolution of the GUT delta without specifying completely
the Yukawa matrices. In this situation, it is difficult to correlate
quark and leptonic deltas.  Finally, the effects of neutrino Yukawas
on the relations between charged-lepton A-parameters is expected to be
similarly small even for large enough (top-quark like) Yukawa
couplings.

We conclude noting that, in general, the RG effects do not
significantly modify the GUT relations for off-diagonal elements
already present at $M_{GUT}$. However in the presence of right-handed
neutrinos and large mixings in the neutrino Yukawa couplings, the new
contributions to the off-diagonal entries in the slepton mass matrices
can destroy the quark-lepton correlations and the analysis becomes
model dependent.

\section{Constraints}
\label{sec:Constraints}

In this section, we list the constraints we have imposed on 
the SUSY parameter space before starting the analysis of
flavor physics in the leptonic and hadronic sectors.

\subsection{Direct SUSY search}
\label{sect:susysearch}

The model we use in this work is a minimal departure from the usual
Constrained MSSM (CMSSM) where we add a single flavor changing entry
in the sfermion couplings at $M_{\rm GUT}$. The usual CMSSM contains
(assuming vanishing flavor blind SUSY phases) five parameters:
$M_{1/2}$, $m_0$, $A_0$, $\tan \beta$ and sign($\mu$)).\footnote
{Notice, we are not enforcing here the minimal supergravity relation
  between the $A$ and $B$ parameters, $B= A-1$} In addition we include
the single MI relevant to the process(-es) of interest at the GUT
scale. Here, we take the $\mu$ sign positive as required by the $b\to
s \gamma$ and muon anomalous magnetic moment constraints. Therefore
the SUSY parameters that enter our analysis for a given MI are only
($m_0$,$m_{1/2}$,$A_0$,$\tan \beta$,$\Delta^f_{ij}$).  We scan the
values of these parameters in the following ranges: $M_{1/2} \le
160$~GeV, $m_{0} \le 380$~GeV, $|A_{0}| \le 3 m_{0}$ and
$5<\tan\beta<15$.  The bound on the $A_{0}$ parameters is set to avoid
charge and/or colour breaking minima \cite{CaDimo}. This would
typically correspond to the following highest mass scales: slepton
masses as large as $m_{\tilde{\ell}}~ \approx 400 $GeV and squark
masses as large as $m_{\tilde{q}}~\approx 550$ GeV; both possibly
observable at LHC.

At the low scale, we impose the following constraints on each point:
\begin{itemize}
\item Lower bound on the light and pseudo-scalar Higgs
  masses~\cite{Higgs_ALEPH};
\item The LEP constraints on the lightest chargino and sfermion
masses~\cite{Eidelman:2004wy};
\item The LEP and Tevatron constraints on squark and gluino
masses~\cite{Eidelman:2004wy}.
\end{itemize}
We consider the MSSM with conserved R-parity and thus the lightest neutralino 
(that coincides with the lightest supersymmetric particle - LSP) provides 
an excellent dark matter candidate.

\subsection{$b\rightarrow s\gamma$}
\label{sect:bsg}

$\BR(B\rightarrow X_s \gamma)$ is particularly 
sensitive to possible non-standard contributions and it provides 
a non-trivial constraint on the SUSY mass spectrum given its precise 
experimental determination and the very accurate SM calculation at the 
NNLO~\cite{Misiak:2006zs}.
When witnessing such a light SUSY spectrum  (we remind that we take 
$M_{1/2} \le 160$~GeV, $m_{0} \le 380$~GeV) a legitimate worry
is whether the bounds on $\BR(B\rightarrow X_s \gamma)$ are respected.
In this work we choose $\mu>0$ which implies destructive interference between
chargino and charged Higgs contributions (and is also preferred by 
the $(g-2)_{\mu}$ constraints).
We have explicitly checked that the combined chargino and charged Higgs 
contributions satisfy the $\BR(B\rightarrow X_s \gamma)$ constraints.
For these points, we checked simultaneously that gluino contributions
satisfy by themselves the $\BR(B\rightarrow X_s \gamma)$ constraints.
Then, these gluino contributions set a bound on the $\delta^{q}_{ij}$ MIs.
Notice that, in this way, we are not allowing the possibility of an accidental 
cancellation of charged-Higgs and chargino contributions with gluino ones.

\subsection{The lightest Higgs boson mass}
\label{sect:mhh}

The non-observation of the lightest neutral Higgs boson ($h^0$) at present
colliders is already a stringent constraint on the MSSM parameter 
space~\cite{Higgs_ALEPH}. Even if the $h^0$ mass depends on the 
whole set of MSSM parameters (after the inclusions of loop corrections),
$m_{h^0}$ mainly depends on (and, indeed, grows with) the left-right mixing 
term in the stop mass matrix $\Mtlr = m_t(A_U-\mu/\tan\beta)$, 
on the average stop mass $\msq$ and on $\tan\beta$.
In particular, it is well known that values of $\tan\beta\leq 2$ are 
strongly disfavored. Taking into account that the lower bound on the $h^0$ 
mass changes with the SUSY parameters, we have explicitly checked that the 
experimental limits \cite{Higgs_ALEPH} are fulfilled in our parameter space.

\subsection{$(g-2)_{\mu}$}

The possibility that the anomalous magnetic moment of the muon 
[$a_\mu = (g-2)_{\mu}/2$], which has been measured very precisely 
in the last few years \cite{g_2_exp}, provides a first hint of physics 
beyond the SM has been widely discussed in the recent literature.
Despite substantial progress both on the experimental 
and on the theoretical sides, the situation is not completely
clear yet (see Ref.~\cite{g_2_th} for an updated discussion).

Most recent analyses converge towards a $2\sigma$ 
discrepancy in the $10^{-9}$ range \cite{g_2_th}:
\beq
 \Delta a_{\mu} =  a_{\mu}^{\rm exp} - a_{\mu}^{\rm SM}
\approx (2 \pm 1) \times 10^{-9}~.
\label{eq:amu_exp}
\eeq

The main SUSY contribution to $a^{\rm MSSM}_\mu$ is usually 
provided by the loop exchange of charginos and sneutrinos.
The basic features of the supersymmetric contribution to $a_\mu$ are correctly
reproduced by the following approximate expression:
\beq
\frac{a^{\rm MSSM}_\mu}{ 1 \times 10^{-9}}  
\approx 1.5\left(\frac{\tan\beta }{10} \right) 
\left( \frac{300~\rm GeV}{m_{\tilde \nu}} \right)^2
\left(\frac{\mu M_2}{m^{2}_{\tilde \nu}} \right)~,
\label{eq:g_2}
\eeq
which provides a good approximation to the full one-loop result
\cite{g_2_mw}. 

The most relevant feature of Eqs.~(\ref{eq:g_2}) is that the sign of
$a^{\rm MSSM}_\mu$ is fixed by the sign of the $\mu$ term so that the
$\mu>0$ region is strongly favored.  This is specially true for the
Standard Model prediction which uses the data from $e^+ e^-$
collisions to compute the hadronic vacuum polarization (HVP).  This
predicts a smaller value than the experimental result by about $3
~\sigma$.  In case one uses the $\tau$ data to compute the HVP, the
discrepancy with SM is reduced to about $1~ \sigma$, but it still
favors a positive correction and disfavors strongly a sizable negative
contribution. Thus, taking $\mu>0$, the region of parameter space
considered in this analysis satisfies the constraint of
\eq{eq:amu_exp}.

\subsection{Electroweak Precision Observables (EWPO)}

The good agreement between the SM predictions and the electroweak
precision observables (EWPO) points to a decoupling of new physics 
contributions to these precision observables. 
As we also consider light superpartners,
we need to take into account the tight constraints on the
supersymmetric spectrum emerging from this agreement.

Several recent and thorough analyses for the MSSM are available in the
literature \cite{EWPO}. The most relevant effect 
is due to  the mass splitting of the superpartners, and in
particular of the third generation squarks. Indeed, large splitting between
$\tilde{m}_{b_L}$ and $\tilde{m}_{t_L}$ 
would induce large contributions to the electroweak $\rho$ parameter.
This universal contribution enters the $Z^0$ boson couplings and the 
relation between $M_W$, $G_\mu$ and $\alpha$ and is therefore significantly 
constrained by present data.
In the cases of the $W$-boson mass and of the effective weak
mixing angle $\sin^2 \Theta_{\rm W}^{\rm eff}$, for example, a
doublet of heavy squarks would induce shifts proportional to its 
contribution to $\rho$: $\delta M_W/M_W \approx 0.72  \,\Delta\rho$ 
and $\delta\sin^2\Theta_{\rm W}^{ eff}\approx - 0.33 \,\Delta\rho$.

In our analysis, we have required that
$\Delta^{\tilde{q}}\rho^{(0)}<1.5\times 10^{-3}$ and we have checked
that no relevant constraints arise from EWPO, as it is confirmed by
the thorough analysis (relative to the CMMSM framework) in
Ref.~\cite{EWPO}.

\section{Mass Insertion bounds from hadronic processes}
\label{sec:MIhadronic}

The comparison of several hadronic flavor-changing processes to their
experimental values can be used to bound the MIs in the different
sectors~\cite{gabbiani}-\cite{Ciuchini:2006dx}. In these analyses it
is customary to consider only the dominant contributions due to gluino
exchange which give a good approximation of the full amplitude,
barring accidental cancellations. In the same spirit, the bounds are
usually obtained taking only one non-vanishing MI at a time,
neglecting the interference among MIs. This procedure is justified
\emph{a posteriori} by observing that the MI bounds have typically a strong
hierarchy, making the destructive interference among different MIs
very unlikely.

\begingroup
\begin{table}
\begin{center}
\begin{tabular}{|c|c|c|}
 \hline\hline
Observable & Measurement/Bound & Ref.\\[0.2pt] 
 \hline
\multicolumn{3}{|c|}{Sector 1--2}\\
$\Delta M_K$ & $(0.0$ -- $5.3) \times 10^{-3}$ GeV & \cite{Yao:2006px}\\
$\varepsilon$ & $(2.232\pm 0.007) \times 10^{-3}$ & \cite{Yao:2006px} \\
$\vert(\varepsilon^\prime/\varepsilon)_{\mathrm SUSY}\vert$ & $< 2 \times 10^{-2}$ &
--\\
\hline
\multicolumn{3}{|c|}{Sector 1--3}\\
$\Delta M_{B_d}$ & $(0.507\pm 0.005)$ ps$^{-1}$ & \cite{hfag}\\
$\sin 2\beta$ & $0.675\pm 0.026$ & \cite{hfag}\\
$\cos 2\beta$ & $>-0.4$ & \cite{Bona:2006ah}\\
\hline
\multicolumn{3}{|c|}{Sector 2--3}\\
BR$(b\to (s+d)\gamma)(E_\gamma > 2.0~\mathrm{GeV})$ & $(3.06\pm 0.49) \times 10^{-4
}$ & \cite{Chen:2001fj}\\
BR$(b\to (s+d)\gamma)(E_\gamma > 1.8~\mathrm{GeV})$ & $(3.51\pm 0.43) \times 10^{-4
}$ & \cite{Koppenburg:2004fz}\\
BR$(b\to s\gamma)(E_\gamma > 1.9~{\mathrm GeV})$ & $(3.34\pm 0.18\pm 0.48) \times
10^{-4}$ & \cite{Aubert:2005cu}\\
$A_{CP}(b\to s \gamma)$ & $0.004\pm 0.036$ & \cite{hfag}\\
BR$(b\to s l^+l^-) (0.04~\mathrm{GeV} < q^2 < 1~\mathrm{GeV})$ &
$(11.34\pm 5.96)\times 10^{-7}$ & \cite{Aubert:2004it,Iwasaki:2005sy}\\
BR$(b\to s l^+l^-) (1~\mathrm{GeV} < q^2 < 6~\mathrm{GeV})$ &
$(15.9\pm 4.9)\times 10^{-7}$ & \cite{Aubert:2004it,Iwasaki:2005sy}\\
BR$(b\to s l^+l^-) (14.4~\mathrm{GeV} < q^2 < 25~\mathrm{GeV})$ &
$(4.34\pm 1.15)\times 10^{-7}$ & \cite{Aubert:2004it,Iwasaki:2005sy}\\
$A_{CP}(b\to s l^+l^-)$ & $-0.22\pm 0.26$ & \cite{Yao:2006px}\\
$\Delta M_{B_s}$ & $(17.77\pm 0.12)$ ps$^{-1}$ & \cite{Abulencia:2006ze} \\
\hline\hline
\end{tabular}
\end{center}
\caption{Measurements and bounds used to constrain the hadronic $\delta^d$'s.}
\label{tab:hexp}
\end{table}
\endgroup

The effective Hamiltonians for $\Delta F=1$ and $\Delta F=2$
transitions including gluino contributions computed in the MI
approximation can be found in the literature together with the
formulae of several observables~\cite{gabbiani}. Even the full NLO
calculation is available for the $\Delta F=2$ effective
Hamiltonian~\cite{Ciuchini:1997bw,Ciuchini:2006dw}.

In our study we use the phenomenological constraints collected in
Table~\ref{tab:hexp}. We use the same set of SUSY parameters described
in the previous Section, so that hadronic and leptonic MIs are related
as discussed in Section~\ref{sec:GUTrel}.  In particular:
\begin{list}{}
\item{\em Sector 1--2}~ The measurements of $\Delta M_K$,
  $\varepsilon$ and $\varepsilon^\prime/\varepsilon$ are used to
  constrain the $\left(\delta^d_{12} \right)_{AB}$ with $(A,B)=(L,R)$.
  The first two measurements, $\Delta M_K$ and $\varepsilon$
  respectively bound the real and imaginary part of the product
  $\left(\delta^d_{12}\right) \left(\delta^d_{12}\right)$. In the case
  of $\Delta M_K$, given the uncertainty coming from the long-distance
  contribution, we use the conservative range in Table~\ref{tab:hexp}.
  The measurement of $\varepsilon^\prime/\varepsilon$, on the other
  hand, puts a bound on Im($\delta^d_{12}$). This bound, however, is
  effective in the case of the LR MI only. Notice that, given the
  large hadronic uncertainties in the SM calculation of
  $\varepsilon^\prime/\varepsilon$, we use the very loose bound on the
  SUSY contribution shown in Table~\ref{tab:hexp}. The bounds coming
  from the combined constraints are shown in
  Table~\ref{tab:MIquarks}. Notice that, here and in the other
  sectors, the bound on the RR MI is obtained in the presence of the
  radiatively-induced LL MI given in Eq.~(\ref{dLL}). The product
  $\left(\delta^d_{12}\right)_{LL} \left(\delta^d_{12}\right)_{RR}$
  generates left-right operators that are enhanced both by the QCD
  evolution and by the matrix element (for kaons only). Therefore, the
  bounds on RR MIs are more stringent than the ones on LL MIs.  
\item{\em Sector 1--3}~ The measurements of $\Delta M_{B_d}$ and
  $2\beta$ respectively constrain the modulus and the phase of the
  mixing amplitude bounding the products $\left(\delta^d_{13}\right)
  \left(\delta^d_{13}\right)$. For the sake of simplicity, in
  Table~\ref{tab:MIquarks} we show the bounds on the modulus of
  $\left(\delta^d_{13}\right)$ only.
\item{\em Sector 2--3}~ This sector enjoys the largest number of
  constraints. The recent measurement of $\Delta M_{B_s}$ constrains
  the modulus of the mixing amplitude, thus bounding the products
  $\vert \left(\delta^d_{23}\right) \left(\delta^d_{23}\right)\vert$.
  Additional strong constraints come from $\Delta B=1$ branching
  ratios, such as $b \to s\gamma$ and $b\to s l^+l^-$. Also for this
  sector, we present the bounds on the modulus of
  $\left(\delta^d_{23}\right)$ in Table~\ref{tab:MIquarks}.
\end{list}

All the bounds in Table~\ref{tab:MIquarks} have been obtained using
the NLO expressions for SM contributions and for SUSY where available.
Hadronic matrix elements of $\Delta F=2$ operators are taken from
lattice
calculations~\cite{Becirevic:2001xt,Allton:1998sm,Babich:2006bh,Nakamura:2006eq}. 
The values of the CKM parameters $\bar\rho$ and $\bar\eta$ are taken
from the {\bf UT}{\it fit} analysis in the presence of arbitrary
loop-mediated NP contributions~\cite{Bona:2006sa}. This conservative
choice allows us to decouple the determination of SUSY parameters from
the CKM matrix.  For $b\to s\gamma$ we use NLO expressions with the
value of the charm quark mass suggested by the recent NNLO
calculation~\cite{Misiak:2006zs}. For the chromomagnetic contribution
to $\varepsilon^\prime/\varepsilon$ we have used the matrix element as
estimated in Ref.~\cite{Buras:1999da}. The $95\%$ probability bounds
are computed using the statistical method described in
Refs.~\cite{Ciuchini:2000de,Ciuchini:2002uv}.

Concerning the dependence on the SUSY parameters, the bounds mainly
depend on the gluino mass and on the ``average squark mass''. A mild
dependence on $\tan\beta$ is introduced by the presence of double MIs
$\left(\delta^d_{ij}\right)_{LL} \left(\delta^d_{jj}\right)_{LR}$ in
chromomagnetic operators. This dependence however becomes sizable only
for very large values of $\tan\beta$.

\begin{table}[t]
\begin{center}
\begin{tabular}{|c|c|c|c|c|}
\hline
~~$ij\backslash AB$ & $LL$ & $LR$ & $RL$ & $RR$ \\
\hline
12 & $1.4\times 10^{-2}$ & $9.0\times 10^{-5}$ & $9.0\times 10^{-5}$ & $9.0\times 10^{-3}$ \\
13 & $9.0\times 10^{-2}$ & $1.7\times 10^{-2}$ & $1.7\times 10^{-2}$ & $7.0\times 10^{-2}$ \\
23 & $1.6\times 10^{-1}$ & $4.5\times 10^{-3}$ & $6.0\times 10^{-3}$ & $2.2\times 10^{-1}$ \\
\hline
\end{tabular}
\end{center}
\caption{95\% probability bounds on $\vert \left(\delta^{d}_{ij}
  \right)_{AB}\vert$  
  obtained using the data set described in
  Section~\ref{sec:Constraints}. See the 
  text for details.} 
\label{tab:MIquarks}
\end{table}

 \section{Mass Insertion bounds from leptonic processes}

 \label{sec:leptons}
 In this section, we study the constraints on slepton mass matrices in
 low energy SUSY imposed by several LFV transitions, namely
 $l_i\rightarrow l_j \gamma$, $l_i\rightarrow l_jl_kl_k$ and
 $\mu$--$e$ transitions in nuclei \cite{Paradisi}. The present and 
projected bounds on 
 these processes are summarized in Table \ref{tab:exp}.  
\begingroup
 \begin{table}
 \begin{center}
 \begin{tabular*}{0.8\textwidth}{@{\extracolsep{\fill}}||c|c|c||}
 \hline\hline
Process & Present Bounds & Expected Future Bounds  \\[0.2pt] 
 \hline
 BR($\mu \to e\,\gamma$) & $1.2~ \times~ 10^{-11}$ &
 $\mathcal{O}(10^{-13} - 10^{-14})$ \\
 BR($\mu \to e\,e\,e$) & $1.1~ \times~ 10^{-12}$ &
 $\mathcal{O}(10^{-13} - 10^{-14})$ \\
 BR($\mu \to e$ in Nuclei (Ti)) & $1.1~ \times~ 10^{-12}$ &
$\mathcal{O}(10^{-18})$ \\
 BR($\tau \to e\,\gamma$) & $1.1~ \times~ 10^{-7}$ &
 $\mathcal{O}(10^{-8}) $ \\
 BR($\tau \to e\,e\,e$) & $2.7~ \times~ 10^{-7}$ &
 $\mathcal{O}(10^{-8}) $ \\
 BR($\tau \to e\,\mu\,\mu$) & $2.~ \times~ 10^{-7}$ &
 $\mathcal{O}(10^{-8}) $ \\
 BR($\tau \to \mu\,\gamma$) & $6.8~ \times~ 10^{-8}$ &
 $\mathcal{O}(10^{-8}) $ \\
 BR($\tau \to \mu\, \mu\, \mu$) & $2~ \times~ 10^{-7}$ &
 $\mathcal{O}(10^{-8}) $ \\
 BR($\tau \to \mu\, e\,e$) & $2.4~ \times~ 10^{-7}$ &
 $\mathcal{O}(10^{-8}) $ \\
\hline\hline
 \end{tabular*}
 \end{center}
 \caption{Present and Upcoming experimental limits on various leptonic 
processes at 90\% C.L.}
 \label{tab:exp}
 \end{table}
 \endgroup
 These processes are mediated by chargino and neutralino loops and
 therefore they depend on all the parameters entering chargino and
 neutralino mass matrices.  In order to constrain the leptonic MIs
 $\delta^{ij}$, we will first obtain the spectrum at the weak scale
 for our SU(5) GUT theory as has been mentioned in detail in section
 \ref{sec:Constraints}.  Furthermore, we take all the flavor
 off-diagonal entries in the slepton mass matrices equal to zero
 except for the entry corresponding to the MI we want to bound.  To
 calculate the branching ratios of the different processes, we work in
 the mass eigenstates basis through a full diagonalization of the
 slepton mass matrix.  So, imposing that the contribution of each
 flavor off-diagonal entry to the rates of the above processes does
 not exceed (in absolute value) the experimental bounds, we obtain the
 limits on the $\delta^{ij}$'s, barring accidental cancellations.

 The process that sets the most stringent bounds is the
 $l_{i}\rightarrow l_{j}\gamma$ decay, whose amplitude has the form
 \bea
 T=m_{l_i}\epsilon^{\lambda}\overline{u}_j(p-q)[iq^\nu\sigma_{\lambda\nu}
 (A_{L}P_{L}+A_{R}P_{R})]u_i(p)\,, \eea where $p$ and $q$ are momenta
 of the leptons $l_k$ and of the photon respectively, $P_{R,L}=
 \frac{1}{2}(1 \pm \gamma_5)$ and $A_{L,R}$ are the two possible
 amplitudes entering the process. The lepton mass factor $m_{l_i}$ is
 associated to the chirality flip present in this transition.  In a
 supersymmetric framework, we can implement the chirality flip in
 three ways: in the external fermion line (as in the SM with massive
 neutrinos), at the vertex through a higgsino Yukawa coupling or in
 the internal gaugino line together with a chirality change in the
 sfermion line.  The branching ratio of $l_{i}\rightarrow l_{j}\gamma$
 can be written as
 \bea
 \frac{BR(l_{i}\rightarrow  l_{j}\gamma)}{BR(l_{i}\rightarrow 
l_{j}\nu_i\bar{\nu_j})} = 
 \frac{48\pi^{3}\alpha}{G_{F}^{2}}(|A_L^{ij}|^2+|A_R^{ij}|^2)\,,
  \nonumber
 \eea
 with the SUSY contribution to each amplitude given by the sum of two terms
$A_{L,R}=A_{L,R}^{n}+A_{L,R}^{c}$. Here $A_{L,R}^{n}$ and $A_{L,R}^{c}$ denote the
contributions from the neutralino and chargino loops respectively.

 \begingroup
 \begin{table}
 \begin{center}
 \begin{tabular*}{0.6\textwidth}{@{\extracolsep{\fill}}||c|c|c|c||}
\hline\hline 
Type of $\delta^l_{12}$ &$\mu \to e\, \gamma$ & $\mu \to e\,e\,e$ & $\mu \to e$
conversion in $Ti$ 
 \\[0.2pt] 
 \hline
 LL & $6 \times 10^{-4}$ & $2\times 10^{-3}$ & $2\times 10^{-3}$ \\
 RR & - & $0.09$ & - \\
 LR/RL & $1 \times 10^{-5}$ & $3.5 \times 10^{-5}$ & $3.5 \times 10^{-5}$ \\
\hline\hline
 \end{tabular*}
 \end{center}
 \caption{Bounds on leptonic $\delta^l_{12}$ from various $\mu \to e$ 
 processes. The bounds are obtained by making a scan of $m_0$ and $M_{1/2}$ 
 over the ranges $m_{0}<380$\,\rm{GeV} and $M_{1/2}<$160\,\rm{GeV}
 and varying $\tan\beta$ within $5<\tan\beta<15$.
 The bounds are rather insensitive to the sign of the $\mu$ mass term.
}
 \label{tab:mue}
 \end{table}
 \endgroup
 Even though all our numerical results presented in Tables 
\ref{tab:mue}--\ref{tab:taumu} 
 are obtained performing an exact diagonalization of sfermion and gaugino
 mass matrices, it is more convenient for the discussion to use the
 expressions for the $l_i \rightarrow l_j \gamma$ amplitudes in the MI
 approximation.
 In particular, we treat both the slepton mass matrix
 and the chargino and neutralino mass matrix off-diagonal elements as
 mass insertions.\footnote{This approximation is well justified and
 reproduces the results of the full computation very accurately in a
 large region of the parameter space \cite{Paradisi}.}
 In this approximation, we have the following expressions
 \bea
 \label{MIamplL}
 A^{ij}_{L}&=&\Frac{\alpha_{2}}{4\pi} \frac{
   \left(\delta^l_{ij}
   \right)_{\LL}}{m_{\tilde l}^{2}}
 ~\Bigg[~f_{1n}(a_2)\!+\!f_{1c}(a_2)\!+\!
 \frac{\mu M_{2}\tan\beta}{(M_{2}^2\!-\!\mu^2)}
 \bigg(f_{2n}(a_2,b)\!+\!f_{2c}(a_2,b)\bigg)\\\nonumber
 &&\qquad\quad+ \tan^2\theta_{W}\,
 \bigg(f_{1n}(a_1)+\mu M_{1}\tan\beta\bigg(\frac{f_{3n}(a_1)}{m_{\tilde l}^{2}}+
 \frac{f_{2n}(a_1,b)}{(\mu^2\!-\!M_{1}^2)}\bigg)\!\bigg)\Bigg]\\
 &+& \Frac{\alpha_{1}}{4\pi}~
 \frac{\left(\delta^l_{ij}
   \right)_{\RL}}{m_{\tilde
l}^{2}}~\left(\frac{M_1}{m_{l_i}}\right)~2~f_{2n}(a_1)\,,\nonumber
 \eea
 \bea
 \label{MIamplR}
 A^{ij}_R=\frac{\alpha_{1}}{4\pi}&\!\!\Bigg(\!\!&\frac{ \left(\delta^l_{ij}
   \right)_{\RR}}{m_{\tilde
l}^{2}}
 \left[4f_{1n}(a_1)+\mu M_{1}\tan\beta\left(\frac{f_{3n}(a_1)}{m_{\tilde l}^{2}}-
 \frac{2f_{2n}(a_1,b)}{(\mu^2\!-\!M_{1}^2)}\right)\right]\\
 &+&
 \frac{ \left(\delta^l_{ij}
   \right)_{\LR}}{m_{\tilde l}^{2}}~\left(\frac{M_1}{m_{l_i}}\right)~
 2~f_{2n}(a_1) \Bigg)\,, \nn
 \eea
where $\theta_W$ is the weak mixing angle, $a_{1,2}=M^{2}_{1,2}/\tilde{m}^{2}$,
$b=\mu^2/m_{\tilde l}^{2}$ and $f_{i(c,n)}(x,y)=f_{i(c,n)}(x)-f_{i(c,n)}(y)$.
The loop functions $f_i$ are given as
\begin{eqnarray}
f_{1n}(x) &=& (-17x^3+9x^2+9x-1+6x^2(x+3)\ln x)/(24(1-x)^5)\nonumber, \\
f_{2n}(x) &=& (-5x^2+4x+1+2x(x+2)\ln x)/(4(1-x)^4), \nonumber \\
f_{3n}(x) &=& (1+9x-9x^2-x^3+6x(x+1)\ln x)/(3(1-x)^5), \nonumber \\
f_{1c}(x) &=&  (-x^3-9x^2+9x+1+6x(x+1)\ln x)/(6(1-x)^5), \nonumber \\
f_{2c}(x) &=& (-x^2-4x+5+2(2x+1)\ln x)/(2(1-x)^4)\,.
\end{eqnarray}
 \begingroup
 \begin{table}
 \begin{center}
 \begin{tabular*}{0.6\textwidth}{@{\extracolsep{\fill}}||c|c|c|c||}
\hline\hline 
Type of $\delta^l_{13}$ &$\tau \to e\, \gamma$ & $\tau \to e\,e\,e$ & $\tau \to e \mu
\mu$ \\[0.2pt] 
 \hline
 LL & $0.15$ & $-$ & -\\
 RR & - & - & - \\
 LR/RL & $0.04$ & $0.5$ & - \\
\hline\hline
 \end{tabular*}
 \end{center}
 \caption{Bounds on leptonic $\delta^l_{13}$ from various $\tau \to e$ 
  processes obtained using the same values of SUSY parameters as in Table
\ref{tab:mue}.}
 \label{tab:taue}
 \end{table}
 \endgroup
 \begingroup
 \begin{table}
 \begin{center}
 \begin{tabular*}{0.6\textwidth}{@{\extracolsep{\fill}}||c|c|c|c||}\hline\hline
 Type of $\delta^l_{23}$ &$\tau \to \mu\, \gamma$ & $\tau \to \mu\,\mu\,\mu$ & $\tau
\to \mu\, e\, e$ 
 \\[0.2pt] 
 \hline
 LL & $0.12$ & - & -\\
 RR & - & - & - \\
 LR/RL & $0.03$ & - & 0.5 \\
\hline\hline
 \end{tabular*}
 \end{center}
 \caption{Bounds on leptonic $\delta^l_{23}$ from various $\tau \to \mu$ 
processes  obtained using the same values of SUSY parameters as in Table
\ref{tab:mue}.}
 \label{tab:taumu}
 \end{table}
\endgroup
We note that all $ \left(\delta^l_{ij}
   \right)_{\LL}$ contributions with internal
chirality flip are $\tan\beta$-enhanced. On the other hand, the only
term proportional to $ \left(\delta^l_{ij}
   \right)_{\LR}$ arises from pure $\tilde B$
exchange and it is completely independent of $\tan\beta$, as can be
seen from Eqs.~(\ref{MIamplL}) and (\ref{MIamplR}). Therefore the
phenomenological bounds on $ \left(\delta^l_{ij}
   \right)_{\LL}$ depend on $\tan\beta$
to some extent, while those on $ \left(\delta^l_{ij}
   \right)_{\LR}$ do not.  The bounds
on LL and RL MIs are expected to approximately fulfill the relation
 $$
  \left(\delta^l_{ij}
   \right)_{\LR} \simeq\frac{m_i}{\tilde{m}}\tan\beta\,\, \left(\delta^l_{ij}
   \right)_{\LL}\,.
 $$
This is confirmed by our numerical study.

The $\delta^d_{\RR}$ sector requires some care because of the presence
of cancellations among different contributions to the amplitudes in
regions of the parameter space.  The origin of these cancellations is
the destructive interference between the dominant contributions coming
from the $\tilde B$ (with internal chirality flip and a
flavor-conserving LR mass insertion) and $\tilde B \tilde H^{0}$
exchange \cite{Paradisi,masina}.  We can see this in the MI
approximation if we compare the $\tan \beta$ enhanced terms in the
second line of \eq{MIamplL} with the $\tan \beta$ enhanced terms in
\eq{MIamplR}. Here the loop function $f_3(a_1)$ corresponds to the
pure $\tilde B$ contribution while $f_{2n}(a_1,b)$ represents the
$\tilde B \tilde H^{0}$ exchange. These contributions have different
relative signs in \eq{MIamplL} and \eq{MIamplR} due to the opposite
sign in the hypercharge of $SU(2)$ doublets and singlets.  Thus, the
decay $l_i\rightarrow l_j\gamma$ does not allow to put a bound on the
RR sector.  We can still take into account other LFV processes such as
$l_i\rightarrow l_jl_kl_k$ and $\mu$--$e$ in nuclei.  These processes
get contributions not only from penguin diagrams (with both photon and
Z-boson exchange) but also from box diagrams.  Still the contribution
of dipole operators, being also $\tan\beta$-enhanced, is dominant.
Disregarding other contributions, one finds the relations \bea
 \label{relations}
 \frac{Br(l_i\!\rightarrow\!l_jl_kl_k)}{Br(l_i\!\rightarrow\!l_j\gamma)}
 \simeq\! 
 \frac{\alpha_{e}}{3\pi}\left(\!\log\frac{m^2_{l_i}}{m^2_{l_k}}\!-\!3\!\right)
 \,,\nn\\
 Br(\mu - e {\rm\ in \ Ti}) \simeq\,\,\alpha_{e}BR(\mu \rightarrow e \gamma)\,,
 \eea
which clearly shows that $l_i \rightarrow l_j \gamma$ is the strongest
 constraint and gives the more stringent bounds on the different 
 $\delta_{ij}$'s.  As we have mentioned above, however,
 in the case of $\delta^l_{\RR}$ 
the dominant dipole contributions interfere destructively in regions
of parameters, 
so that $Br(l_i\rightarrow l_j\gamma)$ 
is strongly suppressed while $Br(\mu - e\rm{\ in \ nuclei})$ and 
 $Br(l_i\rightarrow l_jl_kl_k)$ are dominated 
by monopole penguin (both $\gamma^*$ and Z-mediated) and box diagrams.
The formulae 
for these contributions can be found in Ref. \cite{hisano}.
However, given that non-dipole contributions are typically much 
smaller than dipole ones outside the cancellation region,
 it follows that the bound on $\delta^l_{\RR}$ from $\mu\to eee$ are
 expected to be less stringent than the one on $\delta^l_{\LL}$ from 
 $\mu\to e \gamma$ by a factor $\sqrt{\alpha/(8\pi)}~1/\tan \beta\simeq
 0.02/\tan\beta$, if the experimental upper bounds on the two BRs were the same.
 This is partly compensated by the fact that the present 
 experimental upper bound on the $BR(\mu\to eee)$ is one order of magnitude smaller
 than that on $BR(\mu\to e\gamma)$, as shown in Tab.~\ref{tab:exp}.  
 On the other hand, the process $BR(\mu - e\rm{\ in \ nuclei})$ suffers
 from cancellations through the interference of dipole and non-dipole 
 amplitudes as well. 
 These cancellations prevent us from getting a bound in the RR sector
 from the $\mu$--$e$ conversion in nuclei now as well as in the 
 future when their experimental sensitivity will be improved.
 However, the $\mu\rightarrow e\gamma$ and 
 $\mu$--$e$ in nuclei amplitudes exhibit cancellations in different regions
 of the parameter space so that the combined use of these two constraints
 produces a competitive or even stronger bound than the one we get 
 from $BR(\mu\rightarrow eee)$ alone \cite{Paradisi}.

 We summarize the different leptonic bounds in tables
 \ref{tab:mue}--\ref{tab:taumu}. All these bounds are obtained making
 a scan of $m_0$ and $M_{1/2}$ over the ranges $m_{0}<$380\,GeV and
 $M_{1/2}<$160\,GeV and therefore correspond to the heaviest possible
 sfermions. As expected, the strongest bounds for $\delta^l_{\LL}$ and
 $\delta^l_{\LR}$ come always from $\mu \to e \gamma$, $\tau \to \mu
 \gamma$ and $\tau \to e \gamma$ processes.  In the case of
 $\delta^l_{\RR}$ we can only obtain a mild bound for
 $\left(\delta^l_{12}\right)_{\RR}$ from $\mu \to e e e$ and there are
 no bounds for $\left(\delta^l_{23}\right)_{\RR}$ and
 $\left(\delta^l_{13}\right)_{\RR}$. Notice, however, that does not
 mean that these LFV processes are not effective to constrain the SUSY
 parameter space in the presence of RR MIs. For most of the values of
 $m_0$ and $M_{1/2}$ there is no cancellation and the values of these
 MI are required to be of the order of the LL bounds. Only for those
 values of $m_0$ and $M_{1/2}$ satisfying the cancellation conditions
 a large value of the RR MI is allowed. Therefore, we must check
 individually these constraints for fixed values of the SUSY
 parameters.

\section{Quark-lepton MI relations in GUT scenarios}
\label{sec:correlations}

In the previous two sections we have collected the MI bounds obtained
from the hadronic and leptonic processes.  In the present section, let
us consider a GUT theory, with the corresponding GUT symmetric
relations holding at the GUT scale.  We will focus on the SU(5) case
as summarized by the relations in Table I.  To make a comparison
between leptonic and hadronic MI bounds we have to take into account
that these bounds have a different dependence on the low-energy
parameters of the theory. On the one hand, the hadronic processes are
dominated by gluino contributions and therefore they mainly depend on
the gluino mass and the ``average squark mass''.\footnote{Note that
  the $\tan\beta$ dependence seeps in once
  we consider the double MIs $\left(\delta^d_{ij}\right)_{LL,RR}
  \left(\delta^d_{jj}\right)_{LR,RL}$.}

On the contrary, as we saw in section \ref{sec:leptons}, the leptonic
bounds depend basically on three parameters: gaugino mass, ``average
slepton mass'' and $\tan \beta$.  In our model squark and slepton
masses originate from a common scalar mass $m_0$ at the GUT scale and
therefore we can relate the average squark and slepton masses.

As we have discussed in section \ref{sec:RGE}, the off-diagonal
elements of sfermion mass matrices are not significantly modified in
the RGE evolution from $M_{\rm GUT}$ to $M_W$. However, under some
conditions, as for example in the presence of large neutrino Yukawa
couplings, the RG evolution can generate sizable
off-diagonal elements in the slepton mass matrices even starting from
a vanishing value at $M_{\rm GUT}$. Clearly these effects are never
present in the squark mass matrices, thus breaking the GUT symmetric
relations. This implies that, given our ignorance on the structure of
neutrino Yukawa couplings, we have to be careful when applying the MI
bounds obtained from quarks to leptons or vice-versa.

In fact, if we obtain a bound on a $\delta^l_{ij}$ MI from a leptonic
process at low scales, we can say that, barring accidental
cancellations, this bound applies both to the mass insertions already
present at $M_{\rm GUT}$ and to the mass insertions generated
radiatively between $M_{\rm GUT}$ and $M_{\nu_R}$. Therefore we can
translate this low-scale bound into a bound on the MI at the GUT
scale. This bound applies also to the squark MI at $M_{\rm GUT}$ and
using RGEs we can transport this bound to the electroweak scale.
For example, in $SU(5)$, we find:
\begin{equation}
 \label{ineqdeltas}
 |(\delta^d_{ij})_{\RR}| ~~~\leq~~~ {m_{L}^2 \over m_{d^c}^2}
 |(\delta^l_{ij})_{\LL}|\,.
\end{equation}
The situation is different if we try to translate the bound from quark
to lepton MIs. An hadronic MI bound at low energy leads, after RGE
evolution, to a bound on the corresponding grand-unified MI at 
$M_{\rm GUT}$, applying both to slepton and squark mass matrices.
However, if the neutrino Yukawa couplings have sizable off-diagonal entries,
the RGE running from $M_{\rm GUT}$ to $M_W$ could still generate a new
contribution to the slepton MI that exceeds this GUT bound. Therefore
hadronic bounds cannot be translated to leptons unless we make some
additional assumptions on the neutrino Yukawa matrices.

On general grounds, given that SM contributions in the lepton sector
are absent and that the branching ratios of leptonic processes
constrain only the modulus of the MIs, it turns out that all the MI
bounds arising from the lepton sector are circles in the $\rm{Re}
\left(\delta^d_{ij}\right)_{AB}$--$\rm{Im}\left(\delta^d_{ij}\right)_{AB}$
plane and are centered at the origin.  In some cases, the hadronic
bounds from $B$ physics constraints are too loose and, in principle,
this would allow the presence of MIs larger than one.  The Mass
Insertion approximation cannot be trusted when the bounds on the
$\delta$ approach values $O(1)$.  Therefore, in our analysis we always
consider MI values smaller than one.\footnote{Even in the case we
  consider the possibility of $O(1)$ $\delta$s, the requirement of
  absence of tachyonic scalar masses in the slepton sector, i.e.
  $\left(\delta^l_{3j}\right)_{AB}\leq 1$, provides a bound on the
  squark MIs through the GUT-symmetric relations among leptonic and
  hadronic MIs (see Table~\ref{tb0})}
\begin{figure}
\begin{center}
\includegraphics[width=15cm]{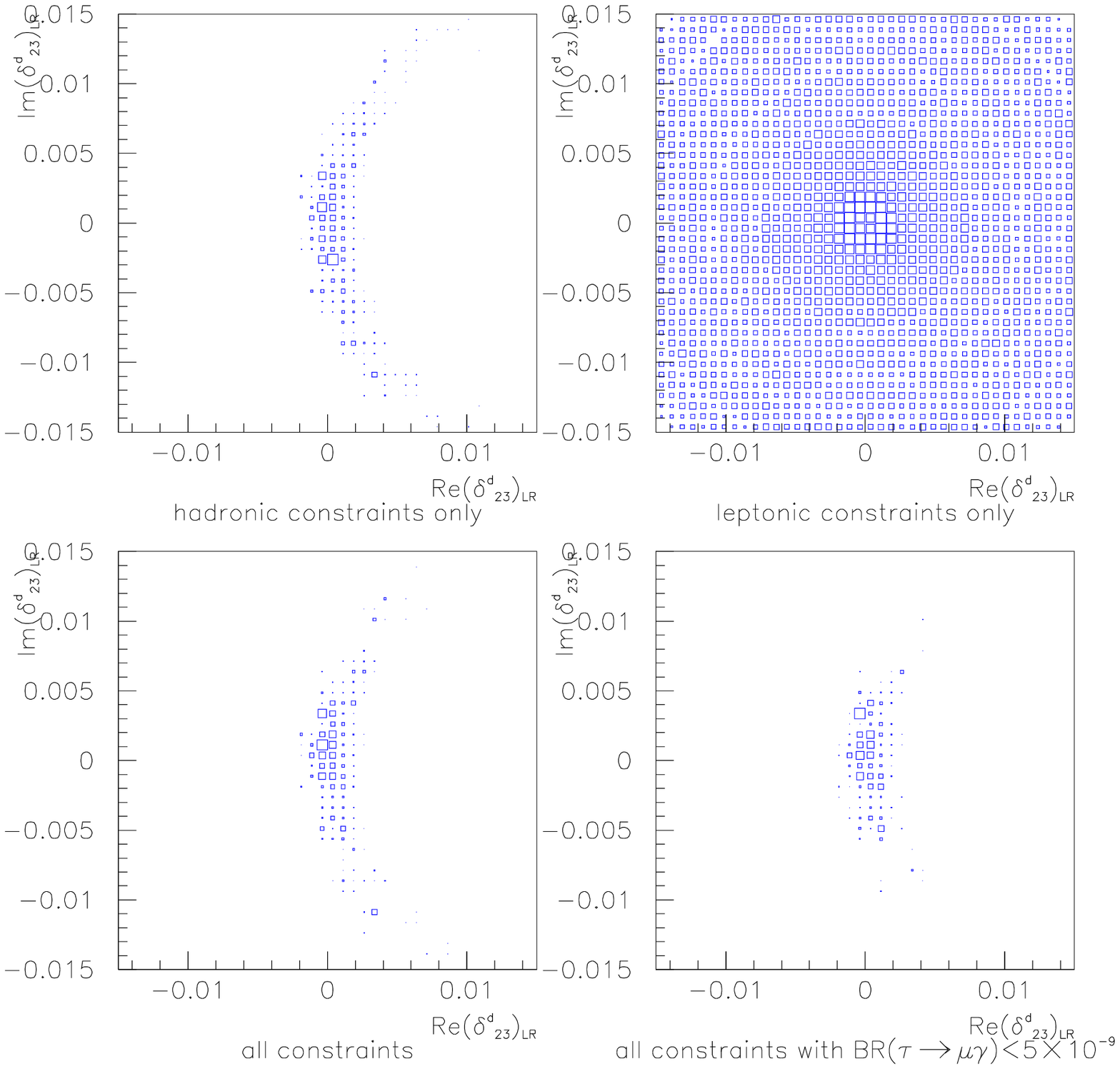}
\end{center}
\caption{Allowed region in the
  Re$\left(\delta^d_{23}\right)_\LR$-Im$\left(\delta^d_{23}\right)_\LR$
  plane using hadronic constraints only (upper left), leptonic
  constraints only (upper right), all constraints (lower left) and all
  constraints with improved leptonic bounds (lower right).}
\label{fig:LR23}
\end{figure}
In the following we will analyze the effect of leptonic bounds on the 
quark mass insertions.
For instance, if we had a $\left(\Delta^{d}_{23}\right)_{\rm LR}$ at
the GUT scale, this would have effects both in the $\tau \to \mu
\gamma$ and $b \to s \gamma$ decays.  Neglecting the effects of
neutrino Yukawas that, if present, could generate an additional
$\left(\delta^{l}_{23}\right)_{\rm LR}$ in the RGE evolution, and
using $\left(\delta^{d}_{23}\right)_{\rm LR} \simeq
(m_b/m_\tau)~(m_{\tilde l}^2/m_{\tilde q}^2)
\left(\delta^{l}_{23}\right)_{\rm RL}$, a bound on
$\left(\delta^{l}_{23}\right)_{\rm RL}$ from the $\tau \to \mu \gamma$
decay translates into a bound on $\left(\Delta^{d}_{23}\right)_{\rm LR}$
thus, into a bound on the SUSY contributions to ${\rm BR}\left(B\to X_s\gamma\right)$.
Similarly, the bound on $\left(\delta^{d}_{23}\right)_{\rm LR}$ would
translate into an upper bound for the $\tau \to \mu \gamma$ branching ratio.

We present the effect of this GUT correlation in our numerical
analysis in Fig.~\ref{fig:LR23}. In the top row, we show the allowed
region in the Re$\left(\delta^{d}_{23}\right)_{\rm
LR}$-Im$\left(\delta^{d}_{23}\right)_{\rm LR}$ plane (larger boxes
correspond to higher probability densities), using hadronic (left) or
leptonic (right) constraints only. We see that the present leptonic
bounds have no effect on the $\left(\delta^d_{23}\right)_\LR$
couplings. This is due both to the existence of strong hadronic bounds
from $b \to s \gamma$ and CP asymmetries and to the relatively weak
leptonic bounds here. Even assuming a future bound on ${\rm BR}
\left(\tau \to \mu \gamma\right)$ at the level of $10^{-8}$,
attainable at B factories, leptonic bounds would marginally improve
the hadronic constraints. We remind the reader that the LR bounds are
basically independent of $\tan \beta$ and hence this fact does not
change for different $\tan\beta$ values.
\begin{figure}
\begin{center}
\includegraphics[width=15cm]{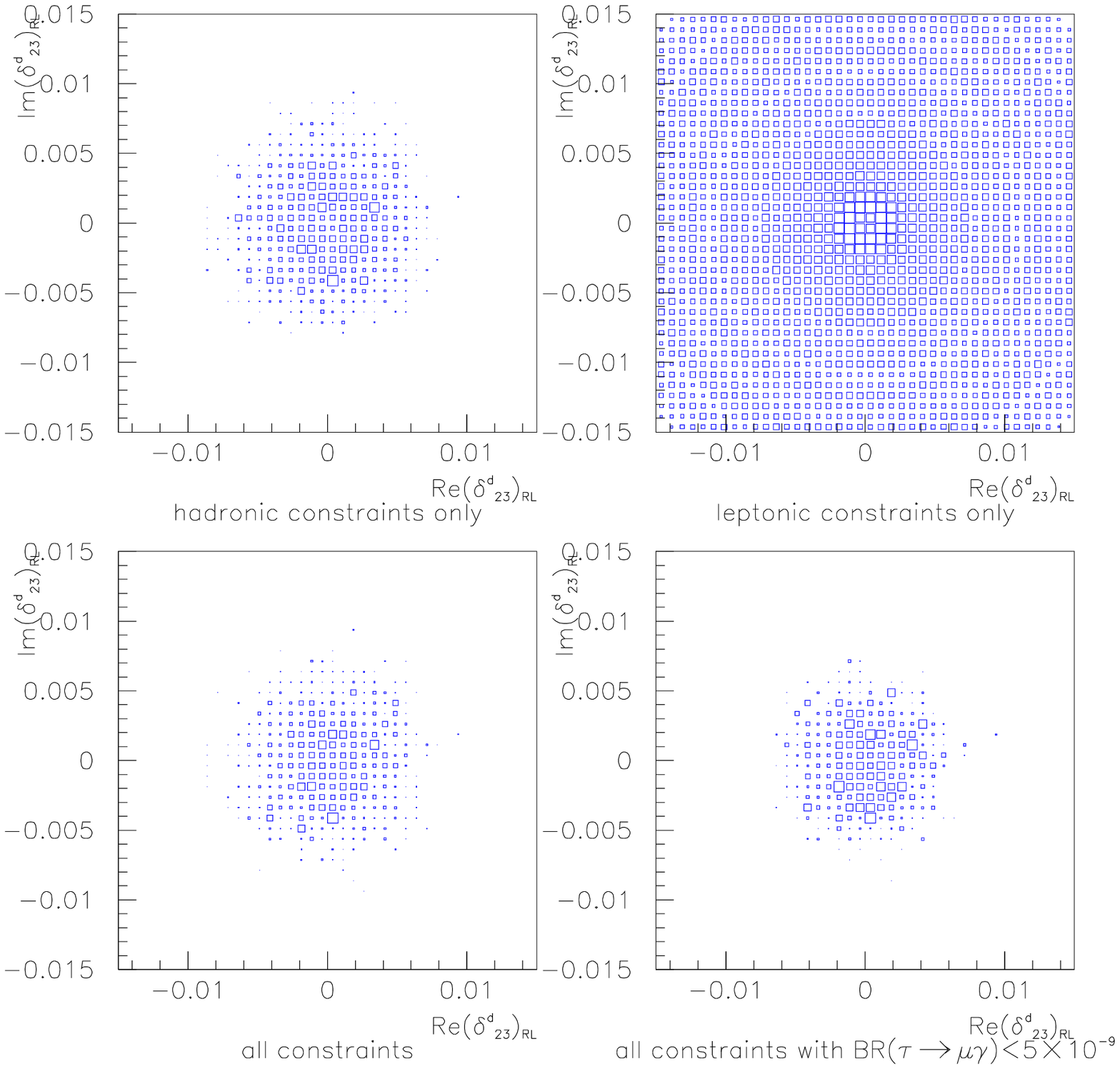}
\end{center}
\caption{Same as Fig.~\protect\ref{fig:LR23} for $\left(\delta^d_{23}\right)_\RL$.}
\label{fig:RL23}
\end{figure}
In Fig.~\ref{fig:RL23} we present the results of the same analysis for  $\left(\delta^d_{23}\right)_\RL$. While the leptonic bounds do not change
with respect to the previous case, the hadronic ones are different as SM
and SUSY $b\to s\gamma$ and $b\to s \ell^+\ell^-$ amplitudes interfere in
the LR case only.

Similarly, if we have a $\left(\Delta^{d}_{23}\right)_{\rm RR}$, 
the corresponding MIs at the electroweak scale are
$\left(\delta^{d}_{23}\right)_{\rm RR}$ and
$\left(\delta^{l}_{23}\right)_{\rm LL}$ that contribute to
$\Delta M_{B_s}$ and $\tau \to \mu \gamma$ respectively
(the impact of $\left(\Delta^{d}_{23}\right)_{\rm RR}$ on
$b\to s\gamma$ and $b\to s \ell^+\ell^-$ is not relevant because
of the absence of interference between SUSY and SM contributions).
\begin{figure}
\begin{center}
\includegraphics[width=15cm]{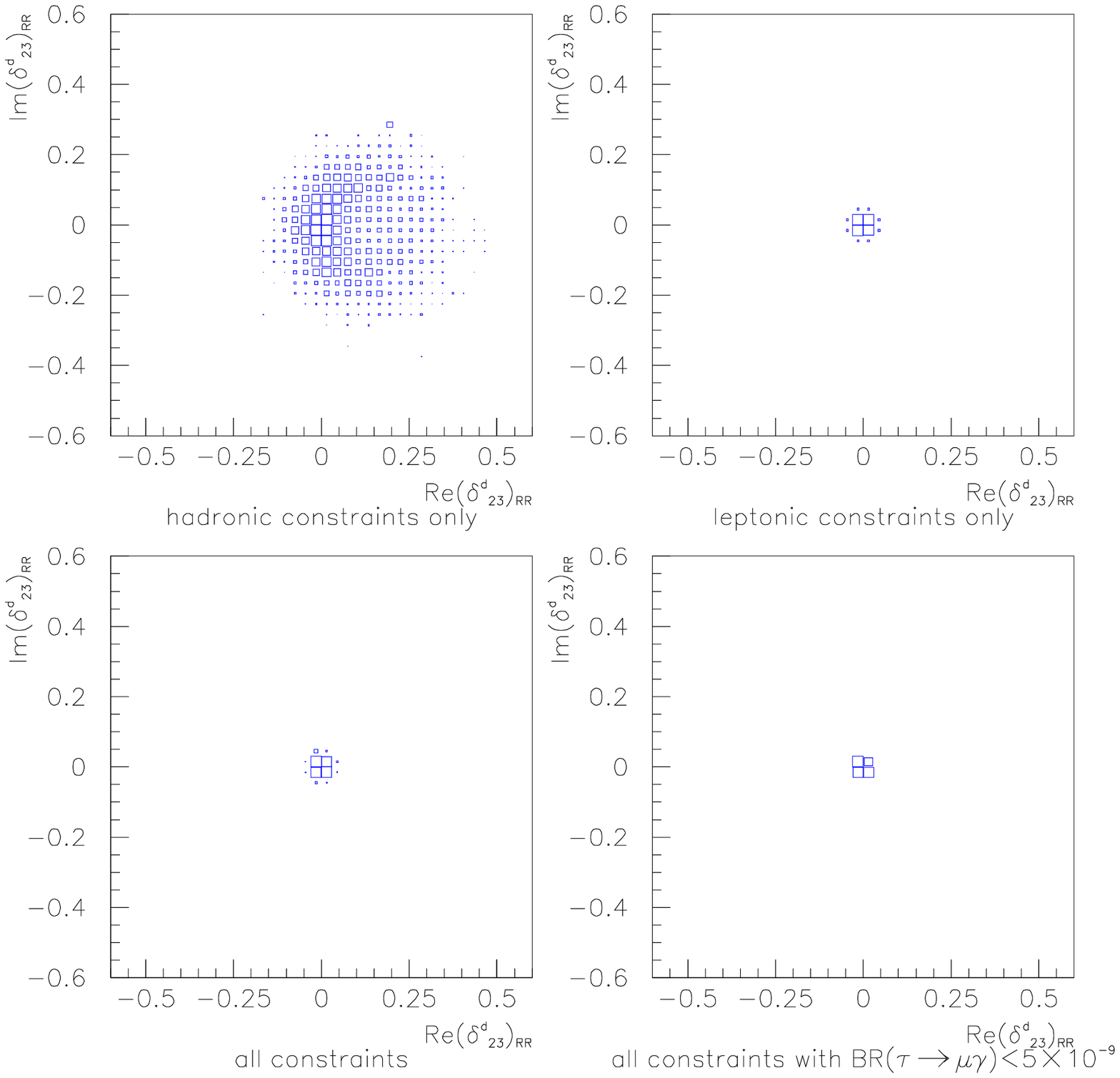}
\end{center}
\caption{Same as Fig.~\protect\ref{fig:LR23} for the $\left(\delta^d_{23}\right)_\RR$.}
\label{fig:RR23}
\end{figure}
In Fig.~\ref{fig:RR23} we present the allowed values of ${\rm
  Re}\left(\delta^d_{23}\right)_\RR$ and ${\rm
  Im}\left(\delta^d_{23}\right)_\RR$ with the different constraints.
The leptonic constraints are quite effective as the bound on the
$BR(\tau\to\mu\gamma)$ from B-factories is already very stringent,
while the recent measurement of $\Delta M_{B_s}$ is less constraining.
The plots correspond to $5<\tan\beta<15$, thus, the absolute bound on
$\left(\delta^l_{23} \right)_{LL}$ is set by $\tan\beta =5$ and it
scales with $\tan\beta$ as $\left(\delta^l_{23}\right)_{LL}\sim
(5/\tan\beta)$.  \footnote{Sizable SUSY contributions to $\Delta
  M_{B_s}$ are still possible from the Higgs sector in the large $\tan
  \beta$ regime both within \cite{buras,lunghi} and also beyond
  \cite{foster} the Minimal Flavor Violating (MFV) framework. However,
  for our parameter space, the above effects are completely
  negligible.}

\begin{figure}
\begin{center}
\includegraphics[width=15cm]{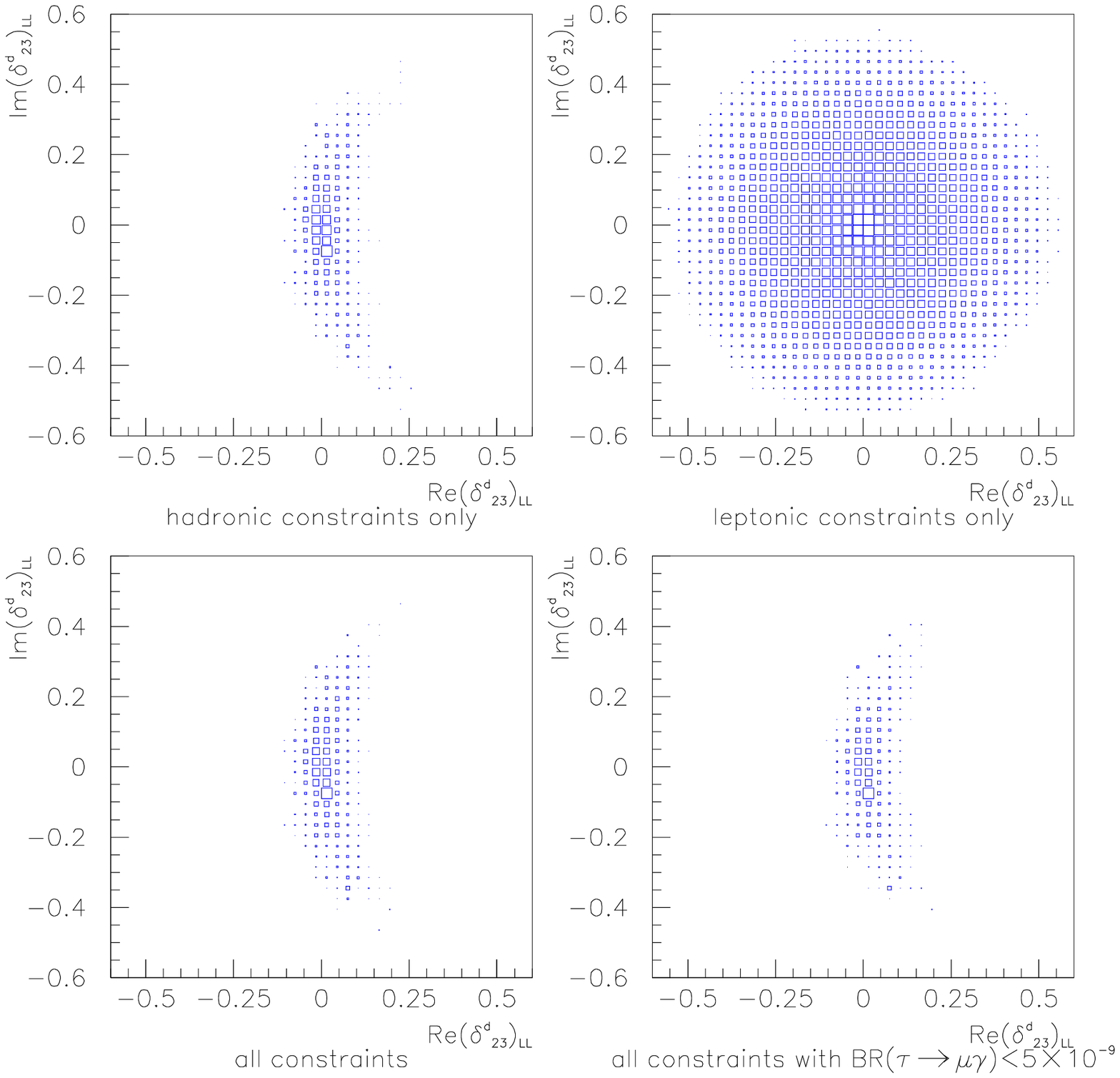}
\end{center}
\caption{Same as Fig.~\protect\ref{fig:LR23} for
  $\left(\delta^d_{23}\right)_\LL$.}
\label{fig:LL23}
\end{figure}

In Fig.~\ref{fig:LL23} we show the results of our analysis for
$\left(\delta^d_{23} \right)_\LL$. In this case, there is no
appreciable improvement from the inclusion of leptonic constraints.
In fact, we remind that $\tau\to\mu\gamma$ is not effective
to constrain $\left(\delta^{l}_{23}\right)_\RR$, i.e. the leptonic MI 
related to $\left(\delta^d_{23} \right)_\LL$ in our SUSY-GUTs scheme,
in large portions of the parameter space because of strong cancellations 
among amplitudes.

\begin{figure}
\begin{center}
\includegraphics[width=15cm]{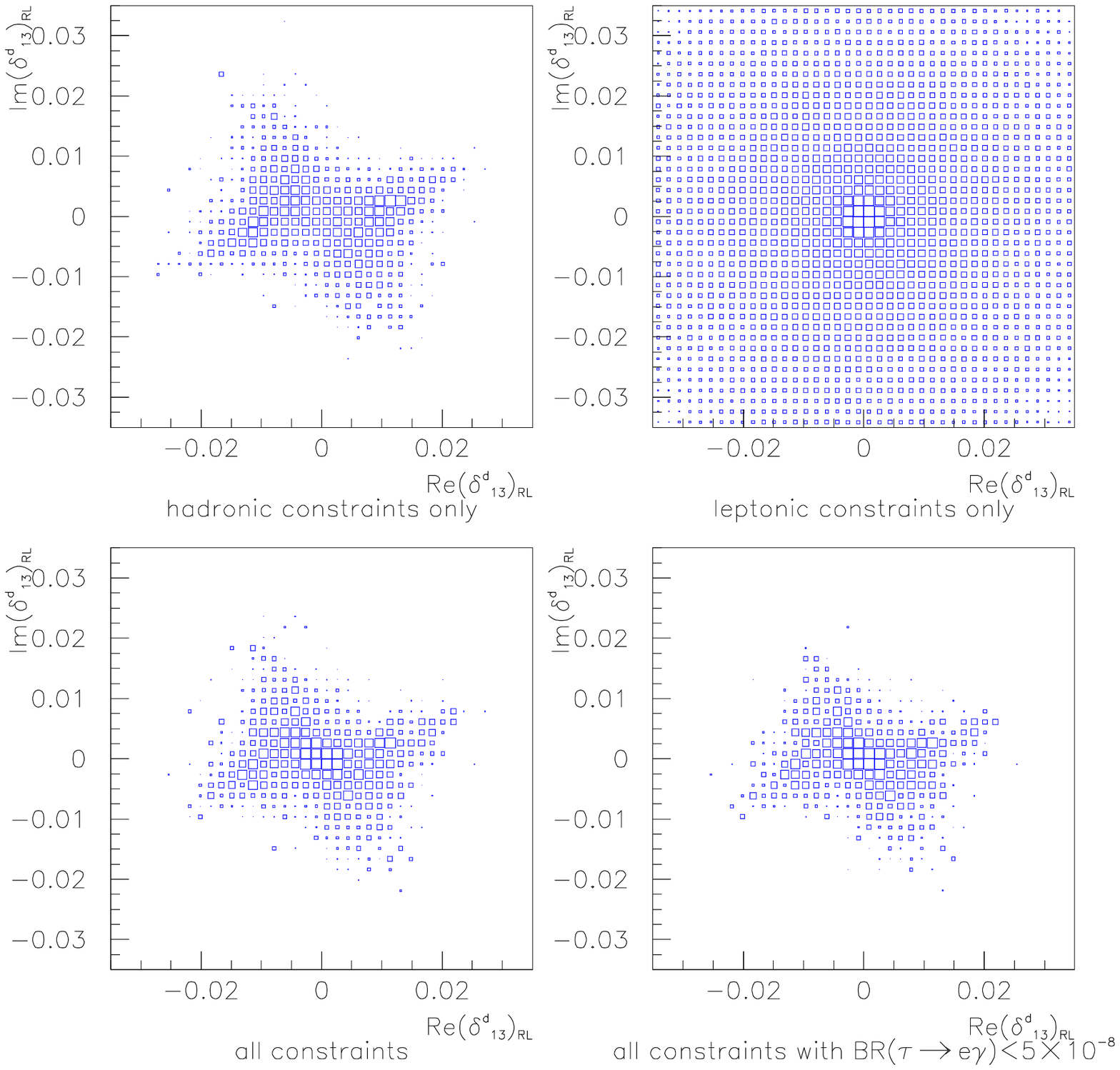}
\end{center}
\caption{Same as Fig.~\protect\ref{fig:LR23} for 
  $\left(\delta^d_{13}\right)_\RL$.}
\label{fig:RL13}
\end{figure}

\begin{figure}
\begin{center}
\includegraphics[width=15cm]{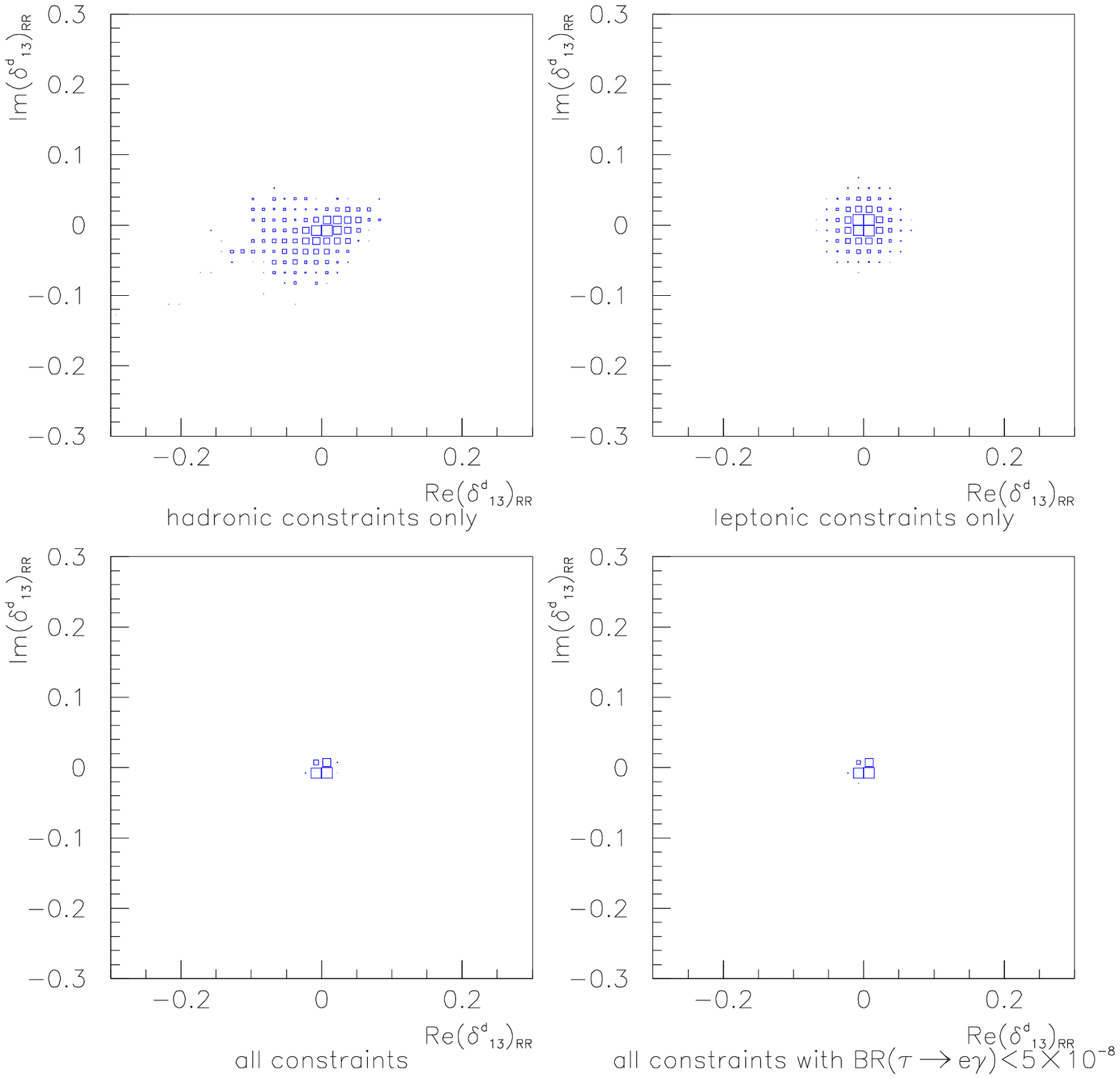}
\end{center}
\caption{Same as Fig.~\protect\ref{fig:LR23} for
  $\left(\delta^d_{13}\right)_\RR$.} 
\label{fig:RR13}
\end{figure}

\begin{figure}
\begin{center}
\includegraphics[width=15cm]{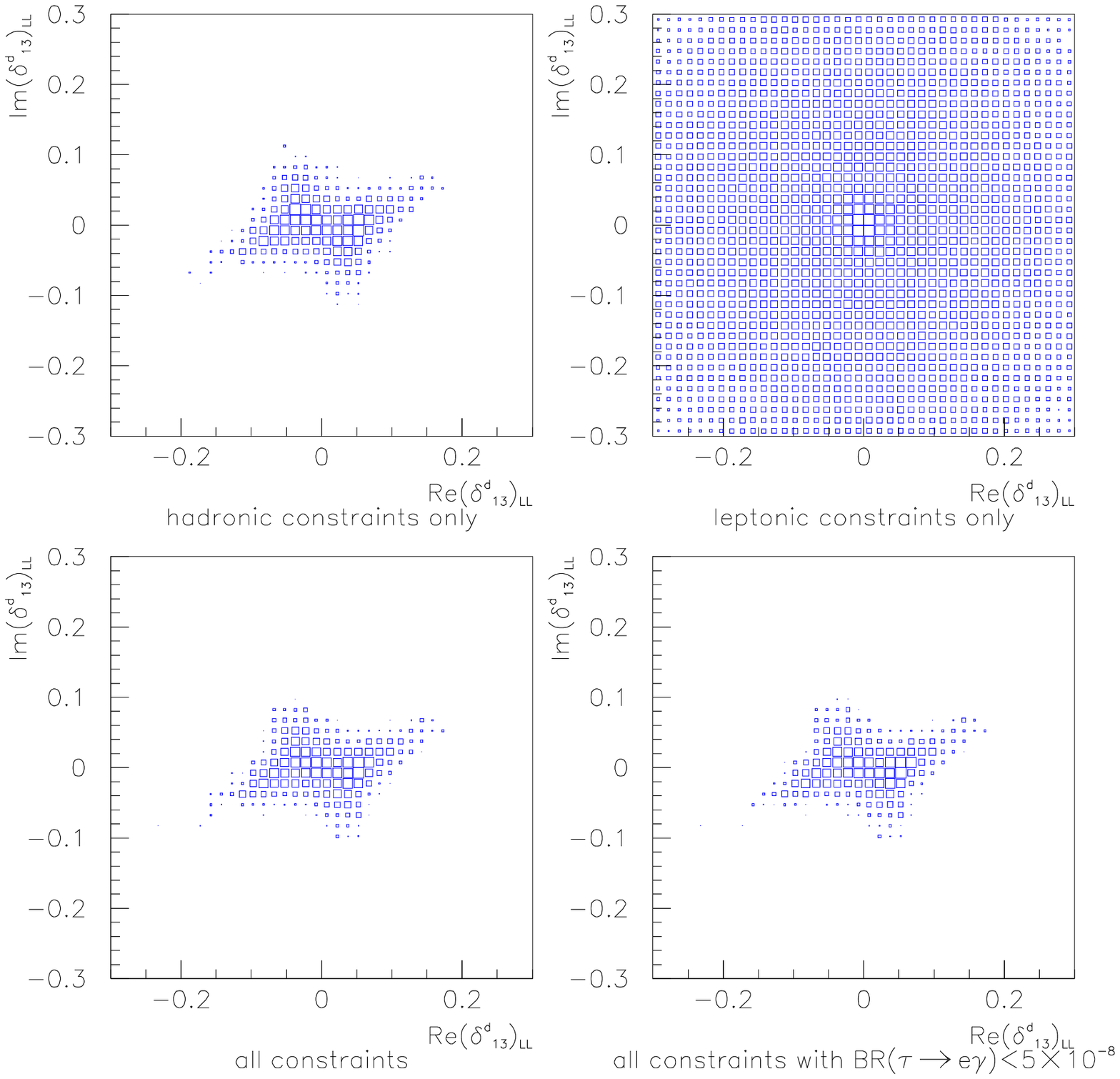}
\end{center}
\caption{Same as Fig.~\protect\ref{fig:LR23} for
  $\left(\delta^d_{13}\right)_\LL$.} 
\label{fig:LL13}
\end{figure}

The analysis of the constraints on the different $\left(\delta^d_{13}
\right)$ MIs is similar to that of the $\left(\delta^d_{23} \right)$ MIs.
In this case, the hadronic constraints come mainly from $\Delta M_{B_d}$ 
and the different CP asymmetries measured at B-factories.
The leptonic bounds are due to the decay $\tau \to e \gamma$.
We present the numerical results in Figs.~\ref{fig:RL13},
\ref{fig:RR13} and \ref{fig:LL13}.
Notice that, in spite of comparable experimental resolutions on
$\Delta M_{B_s}$ and $\Delta M_{B_d}$, the constraints on
$\left(\delta^d_{13}\right)_\RR$ are stronger than those on
$\left(\delta^d_{23}\right)_\RR$ (see
Figs.~\ref{fig:RR23},\ref{fig:RR13}).  The reason is that, in the $13$
sector, we can make use also of the constraints from $2\beta$, in
addition to those relative to $\Delta M_{B_d}$. Moreover, from
Fig.~\ref{fig:RR13} we see that the constraints arising from a
combined analysis of leptonic and hadronic processes are much more
effective then the bounds obtained from the hadronic and leptonic
processes alone.  This is due to the fact that the maximal allowed
values for the hadronic and leptonic deltas in the upper row of
Fig.~\ref{fig:RR13} correspond to different values of $(m_0,M_{1/2})$
in the two cases.\footnote{From Eq.~(\ref{dllckm}) we see that the
  maximal value of the radiatively induced $\delta^d_\LL$ corresponds
  to small $M_{1/2}$ (small $m_{\tilde q}$) and large $m_0$.  The
  largest allowed value for $\delta^d_\RR$ is set by the minimum value
  of this radiatively induced $\delta^d_\LL$, i.e. large $M_{1/2}$ and
  small $m_0/M_{1/2}$. On the other hand the maximal allowed values
  from the leptonic delta correspond to large $M_{1/2}$ and large
  $m_0$.} So, the different $m_0$ dependence of $\Delta M_{B_{d}}$
and $\tau\to e\gamma$ provide the explaination of their interplay in
constraining $\left(\delta^d_{13}\right)_\RR$ (as it is clearly shown
in the lower plot on the left of Fig.~\ref{fig:RR13}). On the other
hand, the above interesting interplay is not effective in the $\Delta
M_{B_s}$ case due to the better bound on the decay $\tau\to\mu\gamma$
and the absence of analog $2\beta$ constraints in the $23$ sector.

Moreover, the leptonic bounds do not have a sizable impact in the RL
and LL cases, as clearly shown in Figs.~\ref{fig:RL13} and
\ref{fig:LL13}, respectively.  The LR case is identical to the RL one.

Finally we analyze the 1--2 sector.
\begin{figure}
\begin{center}
\includegraphics[width=15cm]{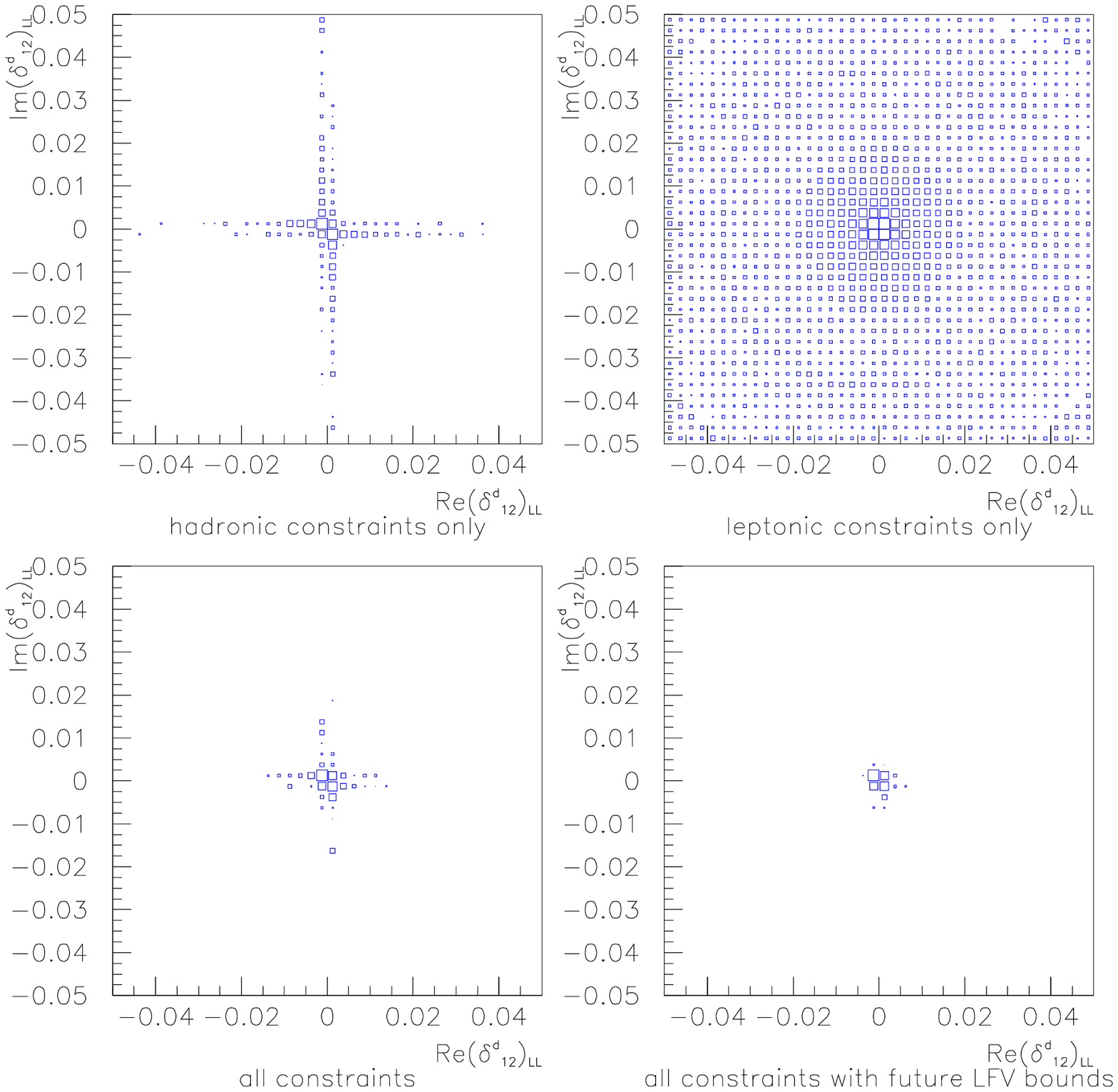}
\end{center}
\caption{Same as Fig.~\protect\ref{fig:LR23} for
  $\left(\delta^d_{12}\right)_\LL$.} 
\label{fig:LL12}
\end{figure}
In Fig.~\ref{fig:LL12} we can see the allowed values of ${\rm
  Re}\left(\delta^d_{12} \right)_\LL$ and ${\rm Im}
\left(\delta^d_{12}\right)_\LL$.  As before, the plot on the left of
the upper row corresponds to the allowed values of these parameters
from hadronic constraints. The dominant hadronic bound comes from
$\varepsilon_K$ which however is ineffective along the ${\rm
  Re}\left(\delta^d_{12} \right)_\LL$ and ${\rm Im}
\left(\delta^d_{12}\right)_\LL$ axes. These directions are eventually
bounded by the milder $\Delta M_K$ constraint. The upper right plot
represents the values allowed taking into account the limits on the
branching ratios of the processes $\mu \to e \gamma$, $\mu \to eee$
and $\mu$--$e$ conversion in nuclei as per the SU(5) relations between
$\left(\delta_{12}^d \right)_\LL$ and $\left(\delta_{12}^l
\right)_\RR$. As we saw in the previous section, the $\mu\to e\gamma$
decay does not provide a bound to this MI due to the presence of
cancellations between different contributions.  We can only obtain a
relatively mild bound, $\left(\delta_{12}^l\right)_\RR\leq 0.09$, if
we take simultaneously into account all the leptonic processes.
However, we see in the lower left plot that, once rescaled by the
factor ${\tilde m_{e^c}^2 \over \tilde m_{d_{\rm L}}^2}$, this bound
is more stringent than $\Delta M_K$, so that it cuts the tails along
the axes. Using the expected bounds for these decays from the proposed
experiments, we obtain the lower right plot in the figure. In these
plots the leptonic constraints come from the monopole and box
contributions to $\mu \to eee$ and $\mu$--$e$ conversion in nuclei and
therefore these bounds are independent of $\tan \beta$. There is a
modest improvement of the bounds on $\left(\delta_{12}^d \right)_\LL$
which however do not take into account possible improvements of
the $\epsilon_K$ constraint.

\begin{figure}
\begin{center}
\includegraphics[width=15cm]{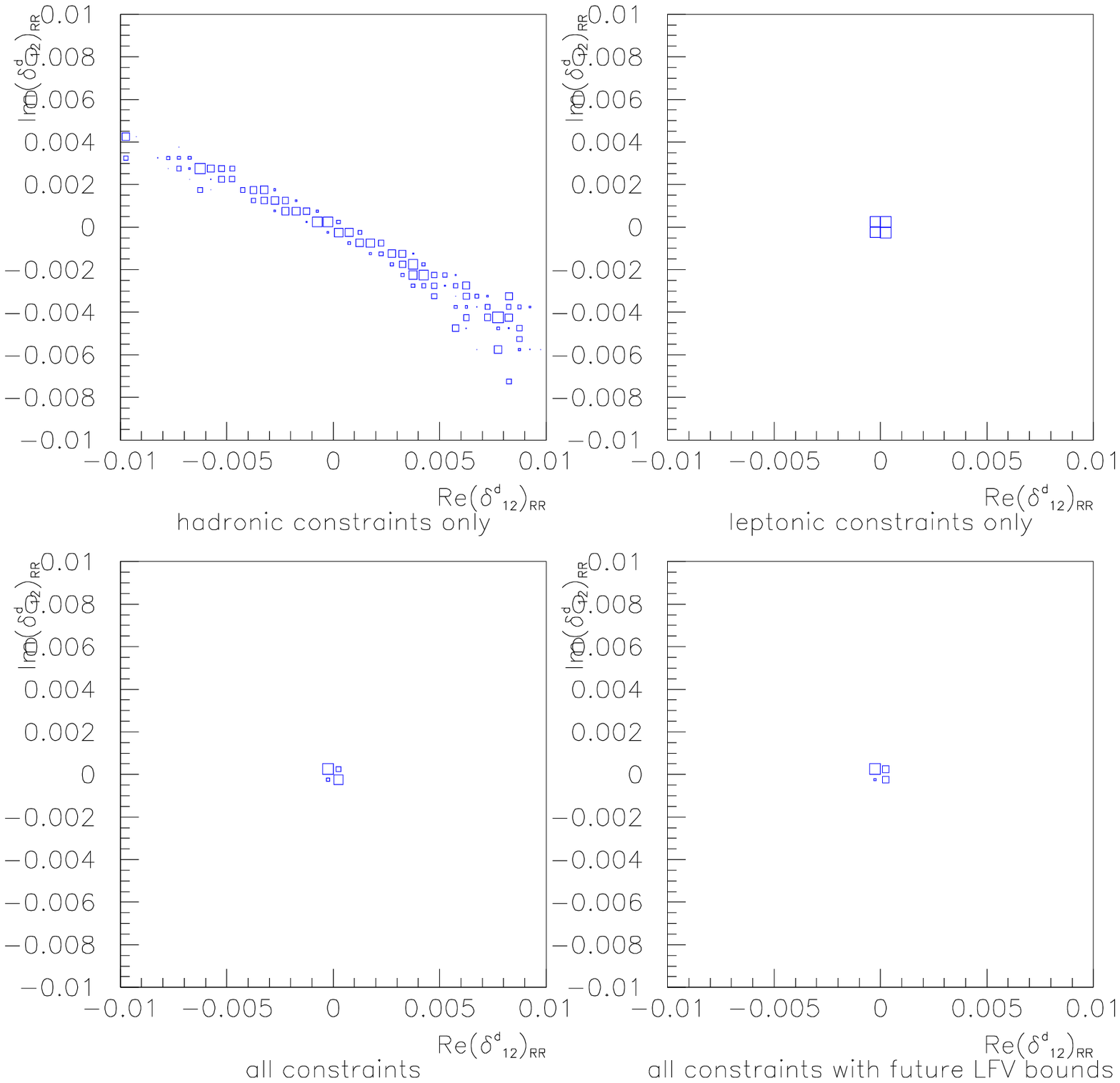}
\end{center}
\caption{Same as Fig.~\protect\ref{fig:LR23} for
  $\left(\delta^d_{12}\right)_\RR$.} 
\label{fig:RR12}
\end{figure}
In Fig.~\ref{fig:RR12} we present the allowed values of ${\rm Re}
\left(\delta^d_{12} \right)_\RR$ and ${\rm Im}
\left(\delta^d_{12}\right)_\RR$. In this case, leptonic constraints,
already using the present upper bound, are competitive and constrain
the direction in which the constraint from $\varepsilon_K$ is not
effective (see the upper left plot). Notice that this direction is
rotated with respect to the LL case because of the presence of LL
$\times$ RR double MIs.

\begin{figure}
\begin{center}
\includegraphics[width=15cm]{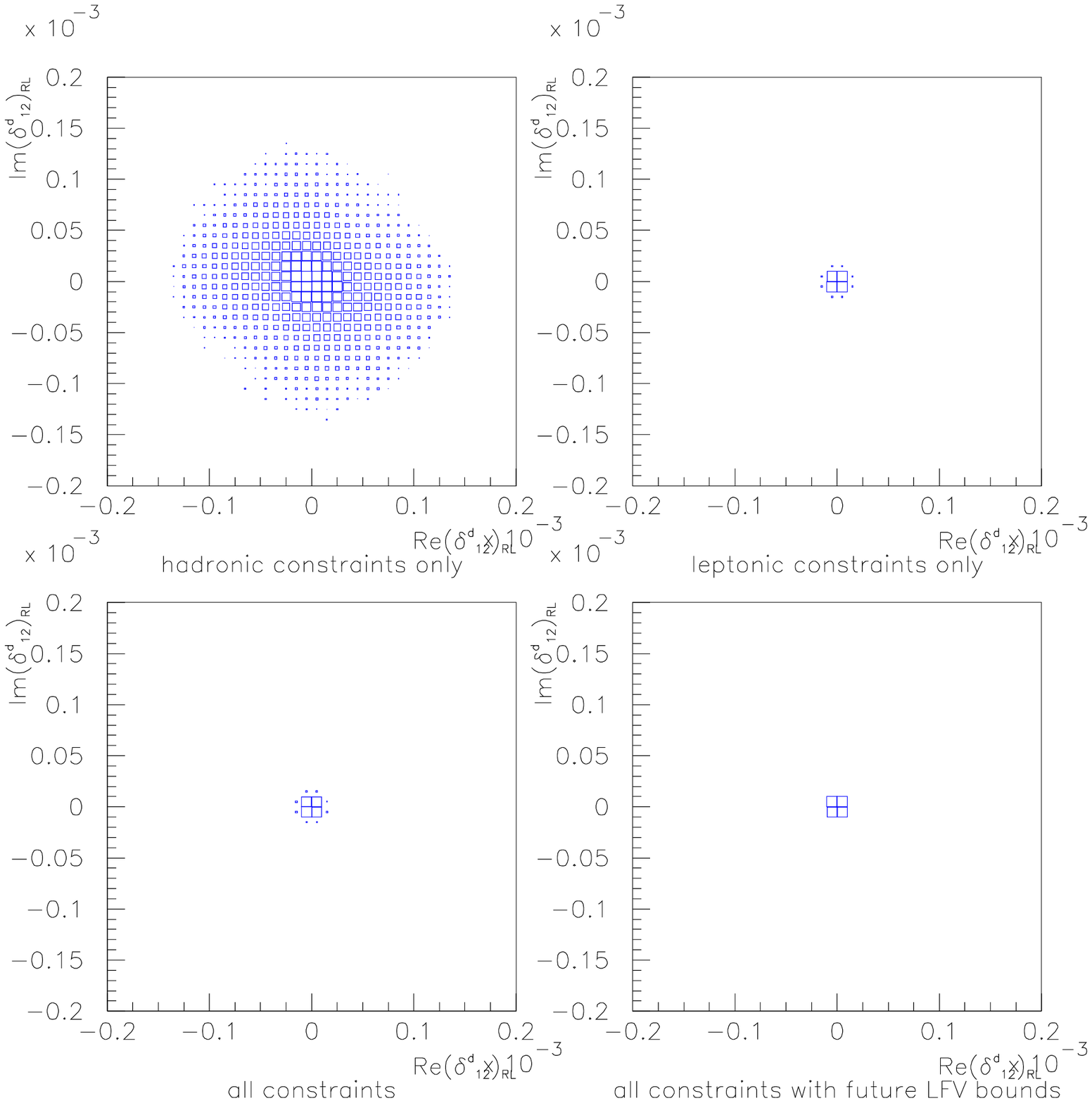}
\end{center}
\caption{Same as Fig.~\protect\ref{fig:LR23} for
  $\left(\delta^d_{12}\right)_\RL$.} 
\label{fig:RL12}
\end{figure}

In Figure ~\ref{fig:RL12} we can see the bounds on ${\rm Re}
\left(\delta^d_{12} \right)_\RL$ and ${\rm Im}
\left(\delta^d_{12}\right)_\RL$. The same bounds apply also to the LR case.
For these MIs, the hadronic bounds come also from $\varepsilon^\prime/\varepsilon$ 
and are quite stringent. However, the bounds from $\mu \to e \gamma$ are even more 
effective. Also in this case the bounds are independent of $\tan \beta$.

\section{Conclusions}
\label{sec:conclusions}

While there exists a huge literature dealing, separately, with FCNC
constraints on the hadronic and leptonic SUSY soft breaking terms,
much less attention has been devoted to the intriguing possibility
that the two sectors may find correlated bounds in SUSY theories with
an underlying grand unified symmetry.

We have pursued such an analysis in the context of a broad class of
theories which are based on two appealing assumptions: i) local SUSY
is broken in the observable sector through gravity mediation with the
corresponding soft breaking terms arising (as momentum-independent
hard terms) at an energy scale close to the Planck mass; ii) the
fundamental gauge symmetry of the theory includes a grand unification
of quarks and leptons which is present down to the typical GUT scale
and hence constrains the form of the supergravity Lagrangian, in
particular its K{\"a}hler potential. The presence of the general
conditions i) and ii) entails some correlation between hadronic and
leptonic soft terms at the superlarge scale where they first arise.

Obviously, the extent to which such correlation survives when
performing the superlarge running of the soft breaking terms from a
scale close to the Planck mass down to the electroweak scale depends
on the new physics present in such long interval. Here we adopted the
two simplest possibilities: just a big desert or a new intermediate
scale, below the GUT scale, where the right-handed neutrinos acquire a
mass in a SUSY see-saw framework. Moreover, to make the problem
treatable in a model independent way, we made two relevant
simplifications on the possible pattern of the soft breaking terms: we
considered that only one of the FC MIs in Eqs.~(\ref{cdeltas1})--(\ref{cdeltas4}) 
is switched on at a time and in the discussion of the hadronic FCNC
constraints we took the gluino exchange to be the representative
source of the SUSY FCNC contributions. On the other hand, in the
leptonic sector where there is no analog of the gluino dominance, we
performed a full computation (in the slepton mass eigenstate basis)
pointing out the possible cancellations which may arise when the
various SUSY contributions are taken into account.  Indeed, the first
part of the present work provides a renewed, thorough and
comprehensive assessment of the bounds on the hadronic and leptonic
MIs taking into account all the relevant pieces of information on the
FCNC phenomenology we have been accumulating in these last years.
Obviously this constitutes the basis for the subsequent work of
correlating the hadronic and leptonic FC MIs which is the main goal
of the paper.

The extent of the impact of such correlation on the present upper
bounds on the FC $\delta$ parameters has been exemplified in the study
of the role of LFV processes in constraining the hadronic $\delta$
parameters in the down-squark sector. The relevant hadronic processes
(in kaon and beauty physics) which are involved in bounding the
$\left(\delta^d_{ij}\right)_{\rm AB}$ (AB=LL,RR,LR,RL) already provide
rather stringent limits (see Table~\ref{tab:MIquarks}) on most of
them. Yet, the inclusion of the correlated constraints arising from
$l_i \to l_j + \gamma$ proves to be extremely powerful for quite a
number of such $\delta$'s. This is the case for 
$\left(\delta^d_{23}\right)_\RR$ (Fig.~\ref{fig:RR23}) 
as well as for $\left(\delta^d_{12}\right)_\RL$ and $\left(\delta^d_{12}\right)_\LR$ (Fig.~\ref{fig:RL12}). 

Leptonic bounds are competitive for $\left(\delta^d_{12}\right)_\LL$ 
(Fig.~\ref{fig:LL12}), $\left(\delta^d_{12}\right)_\RR$ (Fig.~\ref{fig:RR12})
{ and $\left(\delta^d_{13}\right)_\RR$ (Fig.~\ref{fig:RR13})}.
Interestingly enough, this holds true even when we consider the
present upper bounds on the relevant LFV processes and, at least in
some cases, it becomes dramatic when we take into account the future
experimental sensitivities to LFV (namely, the third column of
Table~\ref{tab:exp}).  On the other hand, most of
$\left(\delta^d_{ij}\right)$ FC insertions are essentially
unscathed by the inclusion of the related bounds from LFV. This is
the case for LL, LR and RL MIs in the $13$ and $23$ sectors.

\begin{table}[t]
 \centering
\begin{tabular}{|l|c|}
\hline
\multicolumn{2}{|c|}{quarks vs leptons}\\
\hline
$\left(\delta^d_{12}\right)_{\rm LL}$ & X\\
$\left(\delta^d_{12}\right)_{\rm RR}$ & X\\
$\left(\delta^d_{12}\right)_{\rm LR}$ & 2\\
$\left(\delta^d_{12}\right)_{\rm RL}$ & 2\\
$\left(\delta^d_{13}\right)_{\rm LL}$ & 1\\
$\left(\delta^d_{13}\right)_{\rm RR}$ & X\\
$\left(\delta^d_{13}\right)_{\rm LR}$ & 1\\
$\left(\delta^d_{13}\right)_{\rm RL}$ & 1\\
$\left(\delta^d_{23}\right)_{\rm LL}$ & 1\\
$\left(\delta^d_{23}\right)_{\rm RR}$ & 2\\
$\left(\delta^d_{23}\right)_{\rm LR}$ & 1\\
$\left(\delta^d_{23}\right)_{\rm RL}$ & 1\\
\hline
\end{tabular}
\caption{The final score of the quarks vs leptons match showing
  the dominance of hadronic bounds.}
\label{tab:score}
\end{table}

The above considerations provide a precious tool in the effort to
disentangle the underlying SUSY theory in case some SUSY particles
should show up in LHC physics. It will be very difficult to have some
``direct'' signal of the presence of a grand unified supergravity (for
instance, through the observation and study of proton decay modes).
Looking at correlated SUSY contributions in hadronic and leptonic FCNC
processes, together with some information on the scale of the soft
breaking sector from LHC, we could have some hints on whether there
exists a grand unified underlying symmetry.  Admittedly, even if we
are particularly lucky (for instance we observe SUSY particles $\it
and$ some LFV processes), this is going to be a long term project
which requires a lot of sweat and educated guesses. But, at least, our
paper indicates possible paths to follow to achieve some result in the
difficult task of ``reconstructing'' the correct fundamental SUSY
theory.

In this sense, our work is yet another relevant proof of the
complementarity of flavor and LHC physics in shedding light on such an
underlying new physics beyond the SM.

On a more phenomenological ground, our results can find an important
application in individuating for each FC $\delta$ MI which process
(either hadronic or leptonic) is more suitable to constrain it. If it
is true that in some cases hadronic $\delta$'s find a better limit
when LFV processes are taken into account (as we discussed above),
also the reverse turns out to hold in several circumstances. For
instance, we pointed out that there are cases when no bound emerges
from LFV for some leptonic $\delta$. This happens for
$\left(\delta^l_{23} \right)_\RR$ (see Table~\ref{tab:taumu}) and in
this case we have to make use of FCNC in B physics to extract a bound
on such leptonic FC quantity. There is a healthy competition between
hadronic and leptonic FCNC physics in limiting the SUSY MIs. A
comprehensive score of such hadron versus lepton ``match'' is provided
in Table~\ref{tab:score} which shows the final ranking: quarks win with 
21 points and leptons follow with 12 points.\footnote{Scores in 
Table~\ref{tab:score} and points assignment follow the rules of the Italian 
(world champion) football league.}

In conclusion, we hope that this work may display the richness which
is present in flavor physics once we assume a grand unified
supergravity framework with gravity mediated SUSY breaking. It could
be that, at the end, flavor physics is one of the very few handles we
have to understand from low-energy physics whether Nature has chosen
to possess supersymmetry and grand unification at the root of its
symmetries.

\section*{Acknowledgments}
We acknowledge discussions with E. Dudas and M. Passera.  SKV
acknowledges support from Indo-French Centre for Promotion of Advanced
Research (CEFIPRA) project No: 2904-2 ``Brane World Phenomenology''.
AM thanks the PRIN ``Astroparticle Physics'' of the Italian Ministry
MIUR and the INFN `Astroparticle Physics' special project.  We also
aknowledge support from RTN European programs MRTN-CT-2004-503369
``The Quest for Unification'', MRTN-CT-2006-035482 ``FLAVIAnet'',
MRTN-CT-2006-035863 ``UniverseNet'' and MRTN-CT-2006-035505 ``Heptools''.
O.V. acknowledges partial support from the Spanish MCYT FPA2005-01678.
We thank the CERN theory group for hospitality at various stages of work.

\end{document}